\documentclass[a4paper,11pt]{article}
\usepackage[pdftex]{graphicx}
\usepackage[T1]{fontenc}
\usepackage{lmodern}
\usepackage{slashed}
\usepackage[utf8]{inputenc}
\usepackage[english]{babel}
\usepackage{microtype}
\usepackage{cite}
\usepackage{amsmath}
\usepackage{amssymb}
\usepackage{amsfonts}
\usepackage[nottoc,notlot,notlof]{tocbibind}
\usepackage{upgreek}
\usepackage{mathtools}
\numberwithin{equation}{section}
\allowdisplaybreaks
\usepackage[all]{xy}
\usepackage{color}
\usepackage{graphicx}
\graphicspath{{images/}}
\usepackage{geometry}
\usepackage[toc,page]{appendix}
\usepackage{hyperref}
\usepackage[normalem]{ulem}

\newcommand{\diff}{\mathrm{d}}

\newcommand{\ii}{\mathrm{i}}

\usepackage{bm}
\usepackage{ragged2e}
\usepackage{appendix}
\usepackage{slashed}
\usepackage{bbold}
\usepackage{cancel}

\definecolor{blue-violet}{rgb}{0.54, 0.17, 0.89}
\definecolor{PineGreen}{cmyk}{0.92, 0, 0.59, 0.25}
\definecolor{YellowOrange}{cmyk}{0, 0.42, 1, 0}
\definecolor{orange}{rgb}{0.95, 0.5, 0.1}

\usepackage{tikz}

\begin{document}

\begin{flushright}
\today
\end{flushright}

\begin{center}

{\bf\LARGE Twisting D(2,1;\,$\boldsymbol{\alpha}$) Superspace}
\vskip 8mm
{\bf \large L. Andrianopoli$^{1,2,3}$, B.L. Cerchiai$^{2,3,4}$, R. Matrecano$^{1,2}$, \\ R. Noris$^{5}$, L. Ravera$^{1,2}$, M. Trigiante$^{1,2,3}$}
\vskip 5mm
    {\small $^1$ DISAT, Politecnico di Torino, Corso Duca
    degli Abruzzi 24, I-10129 Turin, Italy\\
    $^{2}$  Istituto Nazionale di
    Fisica Nucleare (INFN) Sezione di Torino, Italy
    \\
   $^{3}$ Arnold - Regge Center, Via P. Giuria 1, 10125 Torino, Italy \\
    $^{4}$ IMATI Milano - CNR, Via A. Corti 12, 20133 Milano, Italy\\
    $^{5}$ Central European Institute for Cosmology,
FZU, Na Slovance 1999/2, 182 21 Prague 8, Czech Republic}
\end{center}
\vskip 6mm
\begin{center}
{\small {\bf Abstract}}
\end{center}

We develop a three-dimensional $\mathcal{N}=4$ theory of rigid supersymmetry describing the dynamics of a set of hypermultiplets $(\Lambda^{\alpha\alpha'\dot{\alpha}'}_I,\,\phi^{\alpha A}_I)$ on a curved AdS$_3$ worldvolume background, whose supersymmetry is captured by the supergroup ${\rm D}^2(2,1;\, \boldsymbol{\alpha})$.
To unveil some remarkable features of this model, we perform two twists, involving the SL$(2,\mathbb R)$ factors of the theory.\\
After the first twist, our spacetime Lagrangian exhibits a Chern-Simons term associated with an odd one-form field, together with a fermionic ``gauge-fixing'', in the spirit of the Rozansky-Witten model. The second twist allows to interpret the ${\rm D}^2(2,1;\, \boldsymbol{\alpha})$ setup as a framework capable of describing massive Dirac particles. In particular, this yields a generalisation of the Alvarez-Valenzuela-Zanelli model of ``unconventional supersymmetry''. \\
We comment on specific values of the combination $\alpha+1$, which in our model is related to a sort of gauging in the absence of dynamical gauge fields.

\vskip 6mm
\vfill
\noindent {\small{\it
    E-mail:  \\
{\tt laura.andrianopoli@polito.it}; \\
{\tt bianca.cerchiai@cnr.it}; \\
{\tt riccardo.matrecano@polito.it}; \\
{\tt noris@fzu.cz}; \\
{\tt lucrezia.ravera@polito.it}; \\
{\tt mario.trigiante@polito.it}.}}
   \eject

\numberwithin{equation}{section}
\renewcommand\baselinestretch{1.2}

\tableofcontents


\section{Introduction}
In recent years, much interest has been devoted to $3D$ Chern-Simons (CS) field theories \cite{Witten:1988hf} and their relations with $4D$ supersymmetric theories.\\
More specifically, in a series of papers \cite{Rozansky:1996bq,Gaiotto:2008sd,Kapustin:2009cd,Koh:2009um}, an interesting relation was unveiled between an $\mathcal{N}=4$ supersymmetric CS theory, based on a gauge group $\mathcal{G}$ and coupled to a set of hypermultiplets \cite{Gaiotto:2008sd} and a CS theory based on a supergroup $\mathcal{SG}$ \cite{Kapustin:2009cd, Koh:2009um}, whose maximal bosonic subgroup is $\mathcal{G}$. \\
In this relation, the hypermultiplets of the former theory arise as descendants from the BRST-covariant gauge-fixing of the odd symmetries in $\mathcal{SG}$ \cite{Rozansky:1996bq}. This correspondence entails a topological twist, involving the Euclidean version of the Lorentz group and one of the two $\rm{SU}(2)$ factors in the $\mathcal{N}=4$ R-symmetry group.

In a seemingly unrelated line of research, stemming from works on unconventional supersymmetry \cite{Alvarez:2011gd}, a $3D$ supersymmetric CS theory on an $\rm{OSp}(2|2)$ supergroup was considered, where the only propagating degrees of freedom are encoded in massive spin-$1/2$ fields. We shall refer to this as the AVZ model, where the aforementioned massive spin-$1/2$ fields are related to the gauge connection $\varPsi_{I\mu}$ of the odd gauge symmetries through the following ansatz:
\begin{equation}\label{ansatz}
  \varPsi_{I\mu}=\ii\,\gamma_{{i}}\,{e^i}_\mu\,\chi^{{\tiny{(\rm{AVZ})}}}_I\,,
\end{equation}
$\chi^{{\tiny{ (\rm{AVZ})}}}_I$ being a spin-$1/2$ field and $e^i_\mu$ the dreibein of the  spacetime where the CS Lagrangian is integrated on.\footnote{We shall use the indices $\mu,\nu,\ldots=0,1,2$ as worldvolume spacetime indices  and $i,j,\ldots=0,1,2$ as anholonomic ones.
} 
These theories are naturally defined on a $3D$ spacetime with negative cosmological constant and the  mass of the propagating spin-$1/2$ fields is related to the spacetime torsion.

Eventually a holographic setup was put forward, in which these $3D$ theories can be interpreted as dual to an AdS$_4$ $\mathcal{N}=2$ bulk supergravity \cite{Andrianopoli:2018ymh, Andrianopoli:2019sip, Andrianopoli:2020zbl}, the first step in this direction being to embed the $3D$ model of \cite{Alvarez:2011gd} with unconventional supersymmetry, within an AdS$_3$ supergravity described, \`{a} la Achucarro-Townsend \cite{Achucarro:1987vz}, as a Chern-Simons theory in three dimensions.\\
Finally, in \cite{Andrianopoli:2019sqe} the unconventional supersymmetry of \cite{Alvarez:2011gd} was revisited by applying the analysis of \cite{Kapustin:2009cd} to an Achucarro-Townsend AdS$_3$-supegravity, described as a Chern-Simons theory on an AdS$_3$ supergroup $\mathcal{SG}$. The ansatz (\ref{ansatz}) was related to the gauge-fixing condition of the odd gauge symmetries in the model, carried out in a covariant BRST setting.\newline

The very presence of a cosmological constant in the three-dimensional theories studied in  \cite{Andrianopoli:2018ymh, Andrianopoli:2019sip,Andrianopoli:2019sqe} represents a substantial difference with respect to the models considered in \cite{Rozansky:1996bq, Gaiotto:2008sd, Kapustin:2009cd, Koh:2009um}. As a first consequence, this implies that the two theoretical constructions differ in the chosen topological twist, which involves different real form{s} of the Lorentz and R-symmetry groups. \\
{Another} important issue {related to the cosmological constant} in the construction of \cite{Andrianopoli:2018ymh, Andrianopoli:2019sip, Andrianopoli:2019sqe}, {following the prescription in \cite{Alvarez:2011gd},} is that the worldvolume spin-connection is identified with an ${\rm SL}(2,\mathbb{R})$-gauge connection in $\mathcal{SG}$. Different choices for this identification amount to the presence or not of a non-trivial spacetime torsion on the worldvolume, which naturally has an ${\rm AdS}_3$ geometry. In particular, as shown in \cite{Andrianopoli:2018ymh,Andrianopoli:2019sip}, the condition for the worldvolume theory to be dual to an AdS$_4$ supergravity background requires its vacuum geometry to be of ${\rm AdS}_3$-type {consistently with the characterization of the Chern-Simons theory on $\mathcal{SG}$ as the description \`{a} la Achucarro-Townsend of an  AdS$_3$ supergravity}.\\

To motivate the present work, let us first review the main features of the construction in \cite{Andrianopoli:2019sqe}, which our analysis here is based on.
In that paper, a Chern-Simons theory on an ${\rm AdS}_3$ supergroup of the form $\mathcal{SG}={\rm OSp}(2|2)\times {\rm SL}(2,\mathbb{R})$ was considered and a covariant gauge-fixing of the odd gauge symmetries of $\mathcal{SG}$ was performed, along the lines of \cite{Kapustin:2009cd}. We shall denote the gauge connection associated with these odd gauge symmetries by $\varPsi^{\alpha}_{I\mu}\,\diff x^\mu$, where $I=1,2$ is an ${\rm SO}(2)$ index and $\alpha=1,2$ is an ${\rm SL}(2,\mathbb{R})$ index. \\
The gauge-fixing introduces a dependence of the theory on the  three-dimensional worldvolume metric, where by worldvolume we mean the base space which the Chern-Simons action is integrated on. By the same token, we shall refer to the $\mathcal{SG}$-algebra-valued  Chern-Simons connections as target space fields. \\
The gauge-fixing of the odd gauge symmetry is implemented by a {covariant} BRST procedure which, in turn, implies the introduction of scalar ghost and anti-ghost fields, to be denoted by $\phi_I^\alpha$ and $\bar{\phi}_I^\alpha$, respectively, with the same index structure as the odd-symmetry parameters of $\mathcal{SG}$, but with opposite spin-statistics. As observed in \cite{Rozansky:1996bq}, these fields behave as ordinary scalars naturally parametrizing a hyper-K\"ahler manifold. One can associate the ghost quantum number with an ${\rm SU}(2)$ fundamental representation, labeled by a new index $A=1,2$, so that the ghost/anti-ghost fields can be grouped in a doublet $\phi^{\alpha A}_I$, where { $\phi^{\alpha 1}_I={\rm Re}(\phi_I^\alpha)$ and $\phi^{\alpha 2}_I={\rm Im}({\phi}_I^\alpha)$.\footnote{With respect to \cite{Andrianopoli:2019sqe}, we use here a different ${\rm SU}(2)$-basis to be labeled by the index $A$.} } \\
A peculiarity of $3D$ Chern-Simons theories is the presence, besides the  BRST symmetries generated by {$\mathcal S,\bar{\mathcal S}$}, {of additional ``vector-BRST'' global symmetries \cite{Delduc:1990je,Vilar:1997fg,DelCima:1998ux,Gieres:2000pv,Andrianopoli:2019sqe}, whose generators are denoted {here} by $\mathcal{S}_{{i}},\,\bar{\mathcal{S}}_{{i}}$, ${i}=0,1,2$. Similarly to the scalar ghosts, the BRST generators can be grouped in} ${\rm SU}(2)$-doublets ${\mathcal S^A},\, \mathcal{S}^A_{{i}}$.
The gauge-fixing constraint is implemented by fermionic  Nakanishi-Lautrup fields $\eta^\alpha_I$.\\
Notice that $\varPsi^{\alpha}_{Ii}$ and $\eta^\alpha_I$ transform with respect to the worldvolume Lorentz group ${\rm SL}(2,\mathbb{R})_L$ as triplets and singlets, respectively. Therefore, following \cite{Andrianopoli:2019sqe}, if we identify ${\rm SL}(2,\mathbb{R})_L$ with the diagonal of the two ${\rm SL}(2,\mathbb{R})$ factors in the worldvolume AdS$_3$ isometry group, to be denoted by ${\rm SL}(2,\mathbb{R})_1'\times {\rm SL}(2,\mathbb{R})_2'$, we can arrange these two fields into a single set of Grassmann fields $\Lambda^{\alpha\alpha'\dot{\alpha}'}_I$,
\begin{equation}
    \Lambda^{\alpha \alpha'\dot{\alpha}'}_I= \ii\,\gamma^{i\,\alpha'\dot{\alpha}'}\,\varPsi^{\alpha}_{i\,I}+a\,\epsilon^{\alpha'\dot{\alpha}'}\,\eta^\alpha_I\,,\label{twist10}
\end{equation}
where $\gamma^{i\,\alpha'\dot{\alpha}'}$ and $\epsilon^{\alpha'\dot{\alpha}'}$ are ${\rm SL}(2,\mathbb{R})_L$-invariant tensors on the worldvolume intertwining between the two fundamental representations of ${\rm SL}(2,\mathbb{R})_1'$ and ${\rm SL}(2,\mathbb{R})_2'$, respectively labeled by $\alpha'=1,2$ and $\dot{\alpha}'=1,2$.
The above choice of the worldvolume Lorentz symmetry inside the bosonic symmetry group corresponds, in our setting, to the topological twist performed in \cite{Rozansky:1996bq, Gaiotto:2008sd, Kapustin:2009cd, Koh:2009um}.\\
In light of the above twist, the Chern-Simons BRST operators $\mathcal{S}^A$ and $\mathcal{S}^A_{{i}}$ can be viewed as components of a single operator with index structure $\mathcal{Q}_{\alpha'\dot{\alpha}' A}$, behaving as supercharges. The latter can then be treated as a global supersymmetry of the gauge-fixed worldvolume theory. The parameters of this supersymmetry have index structure $\epsilon^{\alpha'\dot{\alpha}' A}$ and transform in the $\left({\bf 2},\,{\bf 2},\,{\bf 2} \right)$ of the symmetry group ${\rm SL}(2,\mathbb{R})_1'\times {\rm SL}(2,\mathbb{R})_2'\times{\rm SU}(2)$, where the latter factor is the group acting on the ghost-number of the fields. The number of supercharges is eight, corresponding to an $\mathcal{N}=4$ supersymmetry on $D=3$. \\
These considerations and the AdS$_3$ geometry of our background suggest a superspace description based on an AdS$_3$ supergroup whose maximal bosonic subgroup is a suitable real form of ${\rm SL}(2,\mathbb{C})^3$ and the odd generators transform in the product of the bi-spinor representation of each ${\rm SL}(2,\mathbb{C})$ factor. This naturally hints towards the exceptional supergroup ${\rm D}^2(2,1;\, \boldsymbol{\alpha})$, whose maximal bosonic subgroup is indeed ${\rm SL}(2,\mathbb{R})_1'\times {\rm SL}(2,\mathbb{R})_2'\times{\rm SU}(2)$ and whose odd generators $\mathcal{Q}_{\alpha'\dot{\alpha}' A}$ transform in the product $({\bf 2},\,{\bf 2},\,{\bf 2})$ of the fundamental representations of the three factors \cite{Gunaydin:1986fe}.\\
This construction bears important differences with respect to the one in \cite{Kapustin:2009cd}, due to the fact that we work on a worldvolume with an AdS geometry, with isometry group ${\rm SO}(2,2)\sim {\rm SL}(2,\mathbb{R})_1'\times {\rm SL}(2,\mathbb{R})_2'$. \\
Our model has $\mathcal{N}=4$ supersymmetry: on a Minkowski worldvolume, this would normally be associated with an ${\rm SO}(4)$ R-symmetry group commuting with the spacetime symmetry, as in \cite{Kapustin:2009cd}. In our case, instead, one of the ${\rm SU}(2)$ factors of the ${\rm SO}(4)$ R-symmetry group corresponds to {one of the two} ${\rm SL}(2,\mathbb{R})$ factor{s} in the spacetime isometry group. \\
As explained above, the worldvolume Lorentz group ${\rm SL}(2,\mathbb{R})_L$ is the diagonal subgroup of the ${\rm SL}(2,\mathbb{R})_1'\times {\rm SL}(2,\mathbb{R})_2'$ isometry group. As a consequence of this, in our model the spinor index is a composite one, $\alpha'\dot{\alpha}'$, yielding a redundant 4-component description of the spinorial degrees of freedom and reducing the manifest R-symmetry to ${\rm SU}(2)$. We shall refer to these Grassmann-valued fields as \emph{spinorial} fields, independently of their actual ${\rm SL}(2,\mathbb{R})_L$ representation.
With respect to this supersymmetry, the spinors $\Lambda^{\alpha\alpha'\dot{\alpha}'}_I$ and the scalars $\phi^{\alpha A}_I$ belong to a set of hypermultiplets.\\

In the present work we make a preliminary step towards the explicit construction of the gauge-fixed theory defined in \cite{Andrianopoli:2019sqe}, choosing a worldvolume superspace based on the supergroup ${\rm D}^2(2,1;\, \boldsymbol{\alpha})$. 
More specifically, we construct a $D=3$, $\mathcal{N}=4$ model describing the set of hypermultiplets $(\Lambda^{\alpha\alpha'\dot{\alpha}'}_I,\,\phi^{\alpha A}_I)$ on a rigid AdS$_3$ superspace with symmetry group ${\rm D}^2(2,1;\, \boldsymbol{\alpha})$. Supersymmetric models featuring rigid supersymmetry on a curved background were previously investigated in \cite{Festuccia:2011ws,Butter:2011zt,Butter:2012vm}.\footnote{It is worth emphasizing here that the supergroup ${\rm D}^2(2,1;\, \alpha)$, describes, in the present construction, the worldvolume supersymmetry and should not be mistaken with the Achucarro-Townsend AdS$_3$ supergroup. The latter is in general the product of two supergroups, each containing one of the two ${\rm SL}(2,\mathbb{R})$ factors of the AdS$_3$ isometry group. Let us recall that, in our case, the choice of ${\rm D}^2(2,1;\, \alpha)$ as the worldvolume supergroup was forced by the transformation property of the supersymmetry generators under the ${\rm SL}(2,\mathbb{R})'_1\times {\rm SL}(2,\mathbb{R})'_2\times {\rm SU}(2)$ symmetry of the worldvolume theory.}\\
The hypermultiplets $(\Lambda^{\alpha\alpha'\dot{\alpha}'}_I,\,\phi^{\alpha A}_I)$ transform under the flavour symmetry ${\rm SL}(2,\mathbb{R})\times{\rm SO}(2)$ through the indices $\alpha,\, I$, which we introduced here in view of a future generalisation where a gauging of {the flavour symmetries} will be performed. {In the present paper, the hypermultiplet model, when rewritten in terms of the twisted quantities $(\varPsi^\alpha_{iI},\eta^\alpha_I)$ will bear resemblance with the construction of \cite{Rozansky:1996bq}, restricted to a flat hyper-K\"ahler geometry, where only odd symmetry generators show up in the CS theory.}\\
In \cite{Kapustin:2009cd, Andrianopoli:2019sqe} the presence of a non-gauge-fixed bosonic subgroup $\mathcal{G}$ within the gauge supergroup $\mathcal{SG}$ induces, in the model with gauge-fixed odd-symmetry, mass terms for the fermion fields. The latter are a necessary ingredient if we  ultimately wish to derive, within our theoretical setting, a model featuring unconventional supersymmetry {and describing a massive Dirac field}. In the framework we are considering here, the gauge group $\mathcal{G}$ is absent and the fermion masses are related to the non-trivial gauging of the R-symmetry group ${\rm SU}(2)$ within the supergroup ${\rm D}^2(2,1;\, \boldsymbol{\alpha})$. This in turn depends on the parameter $\boldsymbol{\alpha}$ since the ${\rm SU}(2)$ generators enter the anticommutator of two supersymmetry ones multiplied by $\boldsymbol{\alpha}+1$. As we shall see, the case $\boldsymbol{\alpha}+1=0$ is a singular limit, where the structure of the superalgebra changes. In the general case where $\boldsymbol{\alpha}+1\neq 0$, the group ${\rm SU}(2)$ is non-trivially gauged, the gauge coupling coinciding with the same parameter $\boldsymbol{\alpha}+1$. The corresponding gauge fields $A^x$ are part of the worldvolume supergravity sector which is frozen in the rigid limit we are considering. They are, in other words, solutions, together with the supervielbein and the spin-connection, to the Maurer-Cartan equations of the ${\rm D}^2(2,1;\, \boldsymbol{\alpha})$ superalgebra.\footnote{Here and in the following, with an abuse of notation, we use the same symbols to denote the ${\rm D}^2(2,1;\, \boldsymbol{\alpha})$ supergroup and the associated superalgebra.} As a consequence, their field-strengths are proportional, through the gauge coupling constant, to a non-exact co-cycle in the fermionic directions of superspace (see eq. \eqref{BG4} below). By supersymmetry, the non-trivial gauging of ${\rm SU}(2)$ induces, even in the absence of dynamical gauge fields, spin-$1/2$ fermion shift matrices $\mathbb{N}^\alpha_{IA}$ and a mass term for the fermion fields $\Lambda^{\alpha\alpha'\dot{\alpha}'}_I$, which are then all proportional, through the coupling constant, to $\boldsymbol{\alpha}+1$. This gauging is also responsible for a scalar potential, which is in fact a mass term for the scalar fields.
In other words the ${\rm SU}(2)$ group plays, in our construction, a role to some extent analogous to the one played by the gauge group $\mathcal{G}\subset \mathcal{SG}$ in  \cite{Andrianopoli:2019sqe} in determining the masses of the dynamical fields.\\ 
We find a supersymmetric spacetime Lagrangian, whose superspace extension features a \emph{quasi-invariance} under supersymmetry, meaning an invariance up to a total derivative term, which, in our case only affects the fermionic directions. We shall elaborate on this point in Section \ref{lagrangian}.\\
Related to this, the interpretation of the supersymmetries in terms of the BRST symmetry generators $\mathcal{S}^A$ and their vector counterpart $\mathcal{S}_i^A$ is not straightforward for generic values of $\boldsymbol{\alpha}$, since $\mathcal{S}^A$ do not anticommute on fields with non-trivial ghost-charge,\footnote{For instance we have $\mathcal{S}^2\bar{\phi}\propto (\boldsymbol{\alpha}+1)\,\phi$.} thus failing to define a cohomology. We retrieve a direct BRST interpretation of the ${\rm D}^2(2,1;\, \boldsymbol{\alpha})$ supersymmetries, and thus an apparent connection to the  construction of \cite{Andrianopoli:2019sqe} and \cite{Rozansky:1996bq, Kapustin:2009cd}, only for the special singular value $\boldsymbol{\alpha}=-1$ for which ${\rm SU}(2)$ is effectively ungauged and becomes an external automorphism of the superalgebra. In fact, for such value of the parameter, the ${\rm D}^2(2,1;\,\boldsymbol{\alpha})$ algebra reduces to $${\rm D}^2(2,1;\,\boldsymbol{\alpha}=-1)\simeq \mathfrak{sl}(2|2)\oplus \mathfrak{su}(2)\,$$ and the $\mathfrak{su}(2)$ factor becomes an outer automorphism of the $\mathfrak{sl}(2|2)$ algebra.\\
For the construction of the theory, we adopt the geometric approach to supersymmetry and supergravity, see for instance \cite{DAuria:2020guc}, which allows to obtain the superspace Lagrangian and the supersymmetry transformations of the fields. \\ 
The equations of motion derived from the Lagrangian impose the standard Klein-Gordon equation for the scalar fields, with mass term proportional to the AdS radius and a massive Dirac-like equation for the spinor fields $\Lambda^{\alpha\alpha'\dot{\alpha}'}_I$.\\
By performing a first twist, analogous to the one in \cite{Kapustin:2009cd,Andrianopoli:2019sqe}, the spacetime Lagrangian obtained from the $\mathrm{D}^2(2,1;\,\boldsymbol{\alpha})$ supergroup can precisely be rewritten in terms of the quantities $\varPsi_{I{i}}^\alpha$ and $\eta_I^\alpha$, besides the scalars $\phi^{\alpha A}_I$. As a result, we find that it describes a Chern-Simons term in the connection 1-form $\varPsi_{I}^\alpha$ together with a gauge-fixing term defined by the Nakanishi-Lautrup field $\eta_I^\alpha$ plus a a kinetic term for the scalar fields $\phi^{\alpha A}_I$, analogously to the results in \cite{Rozansky:1996bq,Kapustin:2009cd}. 
Inspection of the Dirac equation, in its twisted form, shows that the only massive degrees of freedom in the fermionic sector are encoded in the field $\eta_I^\alpha$.\\
We eventually perform a second twist to make contact with the model of \cite{Alvarez:2011gd}. In this case, we identify one of the two ${\rm SL}(2,\mathbb{R})$ factors in the isometry group of the AdS$_3$ worldvolume with a part of the aforementioned flavour symmetry group. 
This peculiar choice mixes target space and worldvolume indices and allows to decompose the hyperini in terms of new fields $\hat\chi^\alpha_{{i}I}$ and $\chi^\alpha_I$. The spacetime Lagrangian obtained in this way contains the Chern-Simons term for a spin-3/2 field, $\mathring{\chi}_{I{i}}$ and describes the coupling of this field to two propagating spin-1/2 particles $\chi_{1I},\,\chi_{2I}$.\\
The theory is still consistent if we implement one of the constraints needed for the unconventional supersymmetry, that is if we set the spin-$3/2$ component $\mathring{\chi}_{I{i}}$ to zero: the resulting theory then describes a spin-$1/2$ fermion $\chi_{2I}$ satisfying a Dirac equation, whose mass term is proportional again to $(\boldsymbol{\alpha}+1)$ and which sources the spinor $\chi_{1I}$, which can in general be written as a linear combination of $\chi_{2I}$ and a massless spin-$1/2$ fermion. This leads to a generalisation of the ansatz \eqref{ansatz}, in which the spin-$1/2$ field on the right-hand side is $\chi_{1I}$, while $\chi_{2I}$ is proportional to $\eta^\alpha_I$.\\

The paper is organised as follows. In Section \ref{model} we consider the algebraic relations defining the $\mathrm{D}^2(2,1;\,\boldsymbol{\alpha})$ superalgebra,  {its description as an AdS$_3$ superspace and} the  matter content of our theory, together with  the supersymmetry  transformations laws of the fields. \newline
In Section \ref{lagrangian} we derive the supersymmetric Lagrangian, we compute the corresponding field equations and we prove its supersymmetry invariance both in spacetime and superspace. Furthermore, we comment on the hyper-K\"ahler structure underlying the Lagrangian, in view of a possible generalisation of the theory to a curved scalar manifold. \newline
In Section \ref{twist} we perform two twists of the spinor fields in the hypermultiplets. The first one is useful to make contact with the results obtained in \cite{Rozansky:1996bq} and \cite{Kapustin:2009cd}, whereas the second one allows to implement the (AVZ) ansatz \eqref{ansatz}. \newline
We conclude the paper with some final remarks and possible future developments for this research line.

\section{The Model}
\label{model}

\subsection{The \texorpdfstring{$D=3$}{D=3}, \texorpdfstring{$\mathcal{N}=4$}{N=4} background}

As mentioned in the Introduction, our aim is to construct a theory defined on a supersymmetric background, whose symmetry is captured by the superalgebra $\mathrm{D}^2 (2,1;\,\boldsymbol{\alpha})$.
We start by defining the structure of this algebra and of the superspace based on it.

\subsubsection{The \texorpdfstring{$\mathrm{D}^2(2,1;\,\boldsymbol{\alpha})$}{D21a} superalgebra}

$\mathrm{D}(2,1;\,\boldsymbol{\alpha})$ is an exceptional superalgebra whose bosonic subalgebra is $[\mathfrak{sl}(2)]^3$. It is included in the list of  the superalgebras  defining possible super-AdS backgrounds in three spacetime dimensions, as discussed in \cite{Gunaydin:1986fe}.\\
In particular, we are interested in the supergroup referred to as $\mathrm{D}^2(2,1;\,\boldsymbol{\alpha})$ in \cite{Gunaydin:1986fe}. As explained in the above cited paper, in this case the bosonic subgroup can be chosen in the following real form:
\begin{equation}
\mathrm{D}^2(2,1;\,\boldsymbol{\alpha}) \supset \underbrace{\rm{SL}(2,\mathbb{R})'}_{a=1}\times \underbrace{\rm{SL}(2,\mathbb{R})'}_{a=2} \times \underbrace{\rm{SU}{(2)}}_{a=3} \,, \label{d21ab}
\end{equation}
which allows the interpretation of the first two factors as the isometry group of AdS$_3$, whereas the third represents the manifest part of the R-symmetry. For the sake of notational simplicity we shall denote each of the three factors on the right-hand side of \eqref{d21ab}, generically by SL$(2)_{(a)}$. \\
Finally, the generators of the odd part of the $\mathrm{D}^2(2,1;\,\boldsymbol{\alpha})$ superalgebra, as previously anticipated, transform in the $(\mathbf{2},\mathbf{2},\mathbf{2})$ representation. A detailed description of the superalgebra can be found for example in \cite{Frappat:1996pb}.

Let us denote by $\mathcal{T}^{i_a}_{(a)}$ ($i_a=1,2,3$) the generators of the $\mathfrak{sl}(2)_{(a)}\subset \mathrm{D}^2 (2,1;\,\boldsymbol{\alpha})$ and by $\mathcal{Q}_{\alpha_1 \alpha_2 \alpha_3}$ ($\alpha_a=1,2$) the odd ones. \\
The superalgebra is then expressed by the following (anti)commutation relations:
\begin{align}\nonumber
    &\left[\mathcal{T}^{i_a}_{(a)} \,, \mathcal{Q}_{\ldots\alpha_a \ldots} \right] = \left( \mathbb{t}^{i_a}_{(a)} \right)^{\phantom{\alpha_a}\beta_a}_{\alpha_a}  \mathcal{Q}_{\ldots \beta_a \ldots} \,; \\ \nonumber
    &\lbrace{ \mathcal{Q}_{\alpha_1 \alpha_2 \alpha_3} \,, \mathcal{Q}_{\beta_1 \beta_2 \beta_3} \rbrace} = \sum_{\substack{a,b,c=1\\ a \neq b \neq c }}^3 \ii\, s_a \left( \mathbb{t}^{i_a}_{(a)}\right)_{\alpha_a \beta_a} \epsilon_{\alpha_b \beta_b} \epsilon_{\alpha_c \beta_c} \mathcal{T}_{(a) \, i_a}= \\
    & = \ii \Bigg [ s_1 \left(\mathbb{t}^{i_1}_{(1)} \right)_{\alpha_1 \beta_1} \epsilon_{\alpha_2 \beta_2} \epsilon_{\alpha_3 \beta_3} \mathcal{T}_{(1)i_1} + s_2 \left(\mathbb{t}^{i_2}_{(2)} \right)_{\alpha_2 \beta_2} \epsilon_{\alpha_1 \beta_1} \epsilon_{\alpha_3 \beta_3} \mathcal{T}_{(2)i_2} + s_3 \left(\mathbb{t}^{i_3}_{(3)} \right)_{\alpha_3 \beta_3} \epsilon_{\alpha_1 \beta_1} \epsilon_{\alpha_2 \beta_2} \mathcal{T}_{(3)i_3} \Bigg ] \,, \label{algebra}
\end{align}
where
\begin{align}
    & \left(\mathbb{t}^{i_a}_{(a)}\right)_{\alpha_a \beta_a} = \frac{\ii }{2}\,\left(\gamma^{i_a}_{(a)}\right)_{\alpha_a \beta_a}
\end{align}
are  representation matrices that, taking into account the different real forms of the three bosonic factors, are defined as
 \begin{align}
    & \gamma^{i_1}_{(a=1)} = \gamma^{i_2}_{(a=2)} = \left( \sigma_2 \,, \ii \sigma_1 \,, \ii \sigma_3 \right) \,, \quad \eta=\text{diag}(+,-,-) \,, \\
    & \gamma^{i_3}_{(a=3)} = \left(\sigma_{{1}} \,, \sigma_{{2}} \,, \sigma_3 \right) \,, \quad \eta=\text{diag}(+,+,+) \,.
\end{align}
{We refer to the Appendix \ref{AppA} for useful relations involving the gamma matrices.}\\
In particular, the closure of the algebra imposes the following relation between the three real non-vanishing parameters $s_a${:}  $$s_1+s_2+s_3 = 0\,.$$ The cases where one of the $s_a = 0$ are singular limits.\\
Up to the normalization of the odd generators,  the superalgebra is then characterized by a single parameter $\boldsymbol{\alpha}=s_2/s_1$ and the Jacobi identities can be expressed as the condition
$$s_3/s_1=-(\boldsymbol{\alpha}+1)\,.$$
When expressed in terms of $\boldsymbol{\alpha}$, the singular limits correspond to $\boldsymbol{\alpha}$ equal to $ -1,0,\infty$.\\
A dual representation of the above superalgebra is given in terms of the superconnection
\begin{equation}
    \Omega =  \omega_{(a)i_a} \mathcal{T}^{i_a}_{(a)} + \mathcal{\psi}^{\alpha_1 \alpha_2 \alpha_3} \mathcal{Q}_{\alpha_1 \alpha_2 \alpha_3}\,,\label{conn}
\end{equation}
where the bosonic and fermionic Maurer-Cartan 1-forms $\omega_{(a)i_a}$, $\mathcal{\psi}^{\alpha_1 \alpha_2 \alpha_3}$ define the superalgebra in its dual form through the Maurer-Cartan equations
 \begin{align}
\hat{R}^{i_a}_{(a)}&\equiv  \diff \omega^{i_a}_{(a)}  -\frac 12 \epsilon^{i_a j_ak_a}\omega_{ (a)j_a}\omega_{(a)k_a }
+\frac\ii{2}\,s_a\mathcal{\psi}^{\alpha_1\alpha_2\alpha_3}
\left(\mathbb{t}^{i_a}_{(a)}\right)^{\alpha_1\alpha_2\alpha_3}_{\phantom{\alpha_1\alpha_2\alpha_3}
{\beta_1\beta_2\beta_3}}\mathcal{\psi}^{\beta_1\beta_2\beta_3}\,=\,0 \,,
\label{MC1}\\
\hat{\nabla}\mathcal{\psi}^{\alpha_1\alpha_2\alpha_3}&\equiv \diff \mathcal{\psi}^{\alpha_1\alpha_2\alpha_3}+\sum_{a=1}^3\omega_{i_a(a)}\wedge\left(\mathbb{t}^{i_a}_{(a)}\right)^{\alpha_1\alpha_2\alpha_3}_{\phantom{\alpha_1\alpha_2\alpha_3}
{\beta_1\beta_2\beta_3}}\mathcal{\psi}^{\beta_1\beta_2\beta_3}\,=\,0\,.\label{MC2}
\end{align}
The above equations \eqref{MC1}, \eqref{MC2} can be obtained as Euler-Lagrange equations from the following Chern-Simons Lagrangian:
\begin{align}
\mathcal{L}_{\kappa}=\frac{\kappa}{2}\left[\sum_{a=1}^3\frac 1{s_a}\left(\omega_{(a)i_a} \diff
\omega^{i_a}_{(a)}   -\frac 13 \epsilon^{i_a j_ak_a}\omega_{(a)i_a}\omega_{ (a)j_a}\omega_{(a)k_a } \right)- \ii \mathcal{\psi}_{\alpha_1 \alpha_2 \alpha_3}\hat{\nabla}\mathcal{\psi}^{\alpha_1 \alpha_2 \alpha_3}\right] \,, \label{ellekappa}
\end{align}
where $\kappa$ is the level of the CS action.

\subsubsection{\texorpdfstring{$\mathrm{D}^2(2,1;\,\boldsymbol{\alpha})$}{D21;a} superspace description}

In the following, we are going to give a superspace interpretation of the Maurer-Cartan equations \eqref{MC1}, \eqref{MC2}: to this end, we interpret the diagonal subgroup $\mathrm{SL}(2,\mathbb{R})'_D\subset \mathrm{SL}(2,\mathbb{R})'_{1}\times\mathrm{SL}(2,\mathbb{R})'_{2}$ as the Lorentz group of our background super-geometry. Correspondingly, we choose $$\omega^i \equiv \frac 12(\omega^i_{(1)}+\omega^i_{(2)})\,,$$
as the spin-connection and we interpret the 1-form
$$e^i\equiv \frac L{2}(\omega^i_{(1)}-\omega^i_{(2)})\,,$$
as the dreibein, where we have introduced the scale parameter  $L\in \mathbb{R}_+$ with dimension of a length. The dreibein $e^i$, together with
$\Psi= \sqrt{L}\,\mathcal{\psi}$, which is regarded as a gravitino 1-form field, define the supervielbein of our rigid, but curved superspace background.

Note that, with the above choice, only  $\mathrm{SL}(2,\mathbb{R})'_D\subset \mathrm{SL}(2,\mathbb{R})'_{1}\times\mathrm{SL}(2,\mathbb{R})'_{2}$ is a manifest spacetime symmetry and the indices $i_1$, $i_2$ are both identified with the Lorentz spacetime index $i$. 
We then choose to name the indices referring to the first two factors as follows:
$$i_1=i_2=i=0,1,2\,;\quad \alpha_1 = \alpha'=1,2\,;\quad \alpha_2 = \dot\alpha'=1,2\,;\quad \alpha'\dot\alpha'\equiv(\alpha)=1,\cdots ,4\,.$$
Furthermore, it is useful to introduce the $4\times 4$ matrices
\begin{align}
   \left( \mathbb{T}^{i}_{(1)}\right)_{(\alpha)(\beta)}\equiv \left( \mathbb{t}^{i}_{(1)} \right)_{\alpha' \beta'} \otimes \epsilon_{\dot{\alpha}' \dot{\beta}'} \, , \qquad \left( \mathbb{T}^{i}_{(2)}\right)_{(\alpha)(\beta)}\equiv \epsilon_{\alpha' \beta'} \otimes \left( \mathbb{t}^{i}_{(2)} \right)_{\dot{\alpha}' \dot{\beta}'} ,
\end{align}
whose properties are given in Appendix \ref{AppA}, and their linear combinations
\begin{align}\nonumber
    & \mathbb{J}^i = \mathbb{T}^i_{(1)} + \mathbb{T}^i_{(2)} \,, \\ \nonumber
    & \mathbb{K}^i = \mathbb{T}^i_{(1)} - \mathbb{T}^i_{(2)} \,,\\
    &\mathbb{M}^i_\pm=-\frac \ii 2\left( \mathbb{T}^i_{(1)} \pm \boldsymbol{\alpha} \mathbb{T}^i_{(2)}\right) \,,
\end{align}
 playing the role of the gamma matrices in the ordinary superspace.

On the other hand, only the part of the $\mathcal{N}=4$ R-symmetry associated with the group $\rm{SU}(2)$
is manifest and interpreted as internal symmetry group.
To distinguish it from the other two bosonic factors in the superalgebra, we relabel the corresponding indices
$$i_3,j_3,\ldots=x,y,\ldots=1,2,3\,;\quad \alpha_3,\beta_3,\ldots=A,B,\ldots=1,2\,,$$
we redefine the connection $\omega^{i_3}_{(3)}$ as
$$\omega^{i_3}_{(3)}\Rightarrow A^x$$
and the representation matrix as
\begin{equation}
    \left( \mathbb{t}_{(3)x} \right)^A_{\phantom{A} B} \equiv \frac{\ii}{2} \left( \sigma_{{1}} \,, \sigma_{{2}} \,, \sigma_3 \right)^A_{\phantom{A} B} \,.
\end{equation}

In light of the above definitions, the $\mathrm{D}^2 (2,1;\,\boldsymbol{\alpha})$ Maurer-Cartan equations can be written as
\begin{align}
    & R^i \equiv \diff \omega^i - \frac{1}{2} \epsilon^{ijk} \omega_j \wedge \omega_k = \frac{1}{2L^2} \epsilon^{ijk} e_j \wedge e_k + \frac{1}{2L} \left( \mathbb{M}^i_+  \right)_{(\alpha)(\beta)} \Psi^{(\alpha)A} \wedge \Psi^{(\beta) B} \epsilon_{AB} \,,\label{BG1} \\
    & \nabla \Psi^{(\alpha) A} \equiv \mathcal{D} \Psi^{(\alpha) A} + \left(\mathbb{t}^x_{(3)}\right)^A_{\ B}A^x\wedge \Psi^{(\alpha) B} = - \frac{1}{L} \left(\mathbb{K}^i\right)^{(\alpha)}_{\phantom{(\alpha)}(\beta)}e_i\wedge \Psi^{(\beta) A}  \,, \label{BG2} \\
    & \mathcal{D} e^i \equiv \diff e^i - \epsilon^{ijk} \omega_j \wedge e_k = \frac{1}{2} \left(\mathbb{M}^i_{-}\right)_{(\alpha)( \beta)} \Psi^{(\alpha)A} \wedge \Psi^{(\beta) B} \epsilon_{AB} \,,\label{BG3} \\
    & \mathcal{F}^x \equiv \diff A^x- \frac{1}{2} \epsilon^{xyz} A_y\wedge A_z =   \frac{\ii}{2L} (\boldsymbol{\alpha}+1) \left( \mathbb{t}^x_{(3)}\right)_{AB} \Psi^{(\alpha) A} \wedge \Psi^{ (\beta)B}\,\delta_{(\alpha)(\beta)} \,,\label{BG4}
\end{align}
where the Lorentz-covariant derivative $\mathcal{D} \Psi^{(\alpha) A} $ is defined as
$$\mathcal{D} \Psi^{(\alpha) A} = \diff \Psi^{(\alpha) A} + \omega^i\mathbb{J}_i^{(\alpha)}{}_{(\beta)}\Psi^{(\beta) A}\,. $$
In eqs. \eqref{BG1}-\eqref{BG4}, the left-hand sides can be read as definitions of the superspace curvature, gravitino covariant derivative, supertorsion and gauge field-strength respectively, while the right-hand sides define their parametrizations as 2-forms in the superspace spanned by the supervielbein $e^i$, $\Psi^{(\alpha)A}$. In other words, the above relations define our background superspace. In particular, we see that they define a curved AdS$_3$ background, where $L$ is the AdS radius.\\
Note that the quantity $g{\equiv}(\boldsymbol{\alpha}+1)$ plays the role of the coupling constant associated with the ${\rm SU}(2)$ gauge group. {This is better understood by} redefining $A^x=g\,A^{\prime x}${, in which case} the field strengths of the rescaled gauge fields read
\begin{equation}
    \mathcal{F}^{\prime x} \equiv \diff A^{\prime x}- \frac{g}{2} \epsilon^{xyz} A'_y\wedge A'_z =   \frac{\ii}{2L} \left( \mathbb{t}^x_{(3)}\right)_{AB} \Psi^{(\alpha) A} \wedge \Psi^{ (\beta)B}\,\delta_{(\alpha)(\beta)} \,.
\end{equation}
As it is typical of supersymmetric theories, the gauging of an internal symmetry induces, by supersymmetry, additional terms in the supersymmetry transformation laws (\emph{fermion shifts}) of the fermion fields and fermion mass terms, all proportional to the gauge coupling constant $g$ and a scalar potential proportional to $g^2$, independently of the fact that in our model the gauge fields do not propagate, being part of the background. We are going to compute these terms in the following Section.
In the limit $g=(\boldsymbol{\alpha}+1)\rightarrow 0$ the fermion shifts and the fermion mass terms vanish{, together with the scalar potential}. \\
Let us also mention here that, differently from other AdS superalgebras, the $\mathrm{D}^2 (2,1;\,\boldsymbol{\alpha})$ one allows for {various} different contractions, due to its dependence on the two unrelated parameters $L$ and $\boldsymbol{\alpha}$ and it  admits, in particular, the presence of central charges in the contracted structure both on Minkowski and on AdS. One limit consists in sending $L\to \infty$, in which we recover a super-Poincar\'e structure, in the presence of non-abelian gauge fields; {two more contractions involve} the limit $g=(\boldsymbol{\alpha}+1)\rightarrow 0$, performed \textit{{before or} after} having introduced the redefinition $A^{\prime x}$.
{In the former case we end up with an AdS$_3$ background superspace coupled to non-abelian pure-gauge $\rm{SU}(2)$ connections trivially embedded in superspace, while with the latter contraction} the gauge fields become abelian and are associated with central charges of the contracted superalgebra on an AdS$_3$ background. Finally, one more contraction can be considered, in which we first redefine $A^x=\frac{1}{L} \tilde{A}^x$ and consequently take the limit $L\rightarrow \infty$, obtaining a central extension of a super-Poincar\'e structure where the central charges are associated with abelian gauge fields.

Using standard coset geometry techniques applied to the supercoset ${\rm D}^2(2,1;\,\boldsymbol{\alpha})/[{\rm SL}(2.\mathbb{R})_L\times {\rm SU}(2)]$, one can express the supervielbein $e^i$, $\Psi^{(\alpha)A}$, as well as the connection 1-forms, in terms of the differentials $\diff x^{{\mu}}$, $\diff \theta^{(\alpha)A}$. 
To this end, we can define a supercoset representative
\begin{equation}
\mathbb{L}(x,\,\theta)\equiv \mathbb{L}_F(\theta)\cdot \mathbb{L}_B(x)\,,
\end{equation}
where $\mathbb{L}_F(\theta)=\exp(\theta^{(\alpha) A}\,\mathcal{Q}_{(\alpha) A})$ and $\mathbb{L}_B(x)=\exp(t^i(x)\,\mathcal{K}_i)$, $t^i$ being non-linearly related to the spacetime coordinates and $\mathcal{K}_i\equiv \mathcal{T}_{(1)i}-\mathcal{T}_{(2)i}$. For the sake of simplicity, let us collectively denote the generators of the bosonic subalgebra $\mathfrak{sl}(2)_1 \oplus \mathfrak{sl}(2)_2\oplus \mathfrak{su}(2)$ by $\mathrm{B}_{\mathcal{A}}$. The left invariant 1-form reads
\begin{equation}
\Omega(x,\,\theta,\,\diff \theta,\,\diff x)=\mathbb{L}^{-1}\,\diff \mathbb{L}=\mathbb{L}_B^{-1}(x)\,(\mathbb{L}_F^{-1}\,\diff \mathbb{L}_F(\theta,\,\diff \theta))\,\mathbb{L}_B(x)+\mathbb{L}_B^{-1}\,\diff \mathbb{L}_B(x,\,\diff x)\,.\label{LI1F}
\end{equation}
Defining the Lie algebra-valued 1-forms $\Omega_F(\theta,\diff \theta)$ and $\Omega_B(x,\diff x)$ as follows:
\begin{align}
\Omega_F(\theta,\diff \theta)&\equiv \mathbb{L}_F^{-1}\,\diff \mathbb{L}_F(\theta,\,\diff \theta)=\Omega_F(\theta,\diff \theta)^{(\alpha)A}\,\mathcal{Q}_{(\alpha) A}+\Omega_F(\theta,\diff \theta)^{\mathcal{A}}\,\mathrm{B}_{\mathcal{A}}\,,\nonumber\\
\Omega_B(x,\diff x)&=\mathbb{L}_B^{-1}\,\diff \mathbb{L}_B(x,\,\diff x)=\Omega_B(x,\diff x)^{\mathcal{A}}\,\mathrm{B}_{\mathcal{A}}\,,
\end{align}
we can rewrite the 1-forms in (\ref{LI1F}) as 
\begin{align}
\Omega(x,\,\theta,\,\diff \theta,\,\diff x)&=\Omega_F(\theta,\diff \theta)^{(\alpha)A}\mathbb{L}_B(x)_{(\alpha)A}{}^{(\beta) B}\,\mathcal{Q}_{(\beta) B}+\Omega_F(\theta,\diff \theta)^{\mathcal{A}}\mathbb{L}_B(x)_{\mathcal{A}}{}^{\mathcal{B}}\,B_{\mathcal{B}}+\nonumber\\&+\Omega_B(x,\diff x)^{\mathcal{A}}\,\mathrm{B}_{\mathcal{A}}=\frac{e^i}{L}\,\mathcal{K}_i+\frac{1}{\sqrt{L}}\,
\Psi^{(\alpha)A}\,\mathcal{Q}_{(\alpha)A}+\omega^i\, \mathcal{J}_i+A^x\,\mathcal{T}_{(3)\,x}\,,
\end{align} 
where we have denoted by $\mathbb{L}_B(x)_{(\alpha)A}{}^{(\beta) B}$ and  $\mathbb{L}_B(x)_{\mathcal{A}}{}^{\mathcal{B}}$ the matrices representing the adjoint action of $\mathbb{L}_B(x)$ on the supersymmetry generators $\mathcal{Q}_{(\alpha)A}$ and on ${\rm B}_{\mathcal{A}}$ and which can be deduced from the structure constants of the superalgebra. Moreover we defined $\mathcal J^i=\mathcal T_{(1)}^i+\mathcal T^i_{(2)}$. \newline
From the above equation we can read off the supervielbein and connection. In particular we find for $e^i$ and $\Psi^{(\alpha)A}$ the following  general formulae:
\begin{align}
    e^i&=L\,\left(\Omega_F(\theta,\diff \theta)^{\mathcal{A}}\mathbb{L}_B(x)_{\mathcal{A}}{}^{i}+\Omega_B(x,\diff x)^{i}\right)\,,\nonumber\\
  \Psi^{(\beta)B}&=\sqrt{L}\,\Omega_F(\theta,\diff \theta)^{(\alpha)A}\mathbb{L}_B(x)_{(\alpha)A}{}^{(\beta) B}\,,
\end{align}
where the $i$ index in the first equation labels the components along the ${\mathcal{K}}_i$ generators.
We notice that restriction to spacetime is effected by setting $\theta=0,\,\diff \theta=0$, which in turn implies $ \Psi^{(\beta)B}=0$.\footnote{We refrain here from using an explicit matrix representation of the supercoset representative, since it is known not to be needed in order to compute $e^i$, $\Psi^{(\alpha)A}$ in terms of the coordinates and their differentials. These quantities, indeed, only depend on the structure constants of the superalgebra, as it can be explicitly shown by using the general formula (see Theorem 5 of  \cite{TheFormula}):
$$ e^{-X}\cdot \diff\left(e^X\right)= 
\left(\frac{{\bf 1}-e^{-{\rm Ad}_X}}{{\rm Ad}_X}\right)\,\diff X\,,$$
where $X$ is a superalgebra generator, linear function of the superalgebra parameters, ${\rm Ad}_X(Y)\equiv [X,Y]$ and $\frac{{\bf 1}-e^{-{\rm Ad}_X}}{{\rm Ad}_X}\equiv \sum_{k=0}^\infty \frac{(-1)^k}{(k+1)!}\,({\rm Ad}_X)^k$. In order to evaluate the components of $\Omega_F$ along the generators, as polynomials in $\theta$, $\diff\theta$, for instance, we just need to choose $X=\theta\cdot \mathcal{Q}$ in the above formula. Their explicit expression is not needed for the scope of the present investigation.}\\

In the following, we will study the dynamics of a set of hypermultiplets in this curved background.
To be consistent with the rigid superspace interpretation, the supergravity Lagrangian \eqref{ellekappa} must  decouple from the matter sector in the rigid limit. To this aim, we set the parameter $\kappa$ to
$$\kappa=\frac{L}{\ell_P}\,,$$
where we denote by $\ell_P$ the Planck length. In the rigid limit $\ell_P\to 0$,  it is possible to choose $L\gg\ell_P$, so that the supergravity dynamics is fully decoupled from the matter sector.

\subsection{The matter content of the model}

The model describes the rigid supersymmetric background defined above coupled to a set of hypermultiplets labeled by a couple of flavor indices $a\equiv\alpha I=1,\cdots,2n$, namely a set of scalars $\phi^{\alpha A}_I$ and their spin-$1/2$ superpartners $\Lambda^{\alpha (\alpha)}_I$.

In the geometric approach to supersymmetry and supergravity in superspace,\footnote{A recent detailed description of the geometric approach  can be found in \cite{DAuria:2020guc}.}  the first step for identifying the model is to extend the notion of the matter fields to superfields in superspace and to define their covariant derivatives in superspace,
\begin{align}
    & \nabla \phi^{\alpha A}_I \equiv \diff \phi^{\alpha A}_I + A^x \left( \mathbb{t}^x_{(3)} \right)^A_{\phantom{A}B} \phi^{\alpha B}_I \,, \label{phidef}\\
    & \nabla \Lambda^{\alpha (\alpha)}_I \equiv \diff \Lambda^{\alpha(\alpha)}_I + \omega_i \left(\mathbb{J}^{i}\right)^{(\alpha)}_{\phantom{(\alpha)}(\beta)} \Lambda^{\alpha (\beta)}_I \,.
\end{align}
The corresponding  Bianchi identities, which stem from the $\diff^2$-closure, must then hold on-shell in superspace,
\begin{align}
    & \nabla^2 \phi^{\alpha A}_I = \left( \mathbb{t}^x_{(3)} \right)^A_{\phantom{A}B} \mathcal{F}_{x(3)} \phi^{\alpha B}_I \,,\label{BIphi} \\ 
    & \nabla^2 \Lambda^{\alpha (\alpha)}_I = \left( \mathbb{J}_i \right)^{(\alpha)}_{\phantom{(\alpha)} (\beta)} R^i \Lambda^{\alpha (\beta)}_I \,.\label{BIlambda}
\end{align}
Note that the above relations are not identically satisfied in superspace, but amount to constraints that must hold on-shell, when the covariant derivatives above are parametrized as general 1-forms in superspace.
More precisely, their generic parametrization can be expressed as
\begin{align}
\label{parametrizzazioni1}
    & \nabla \phi^{\alpha A}_I = \nabla_i \phi^{\alpha A}_I e^i + \Psi^{(\alpha)A} \Lambda^\alpha_{(\alpha)I} \,, \\
 \label{parametrizzazioni2}   & \nabla \Lambda^{\alpha (\alpha)}_I = \nabla_i \Lambda^{\alpha (\alpha)}_I e^i + \nabla_i \phi^{\alpha A}_I \left(\mathbb{ m}^i\right)^{(\alpha)(\beta)} \Psi_{(\beta)}^B \epsilon_{AB} + \mathbb{N}^\alpha_{IA}\, \Psi^{(\alpha)A}\,.
\end{align}
The consistency constraints \eqref{BIphi}, \eqref{BIlambda}, once imposed on the parametrizations \eqref{parametrizzazioni1}, \eqref{parametrizzazioni2}, imply the field equations for the fermion fields and also determine the auxiliary matrices $\mathbb{m}^i$, $\mathbb{N}^\alpha_{IA}$
\begin{align}
\left( \mathbb{m}^i \right)^{(\alpha)}_{\phantom{(\alpha)} (\beta)}&=  -\frac{\ii}{2} \left( \mathbb{T}^i_{(1)} - \boldsymbol{\alpha} \mathbb{T}^i_{(2)} \right)^{(\alpha)}_{\phantom{(\alpha)} (\beta)}=\left( \mathbb{M}^i_- \right)^{(\alpha)}_{\phantom{(\alpha)} (\beta)}\,\,,\,\,\,\,
    \mathbb{N}^\alpha_{IA} = \frac{\ii(1+\boldsymbol{\alpha})}{4L}\,\epsilon_{AB}\, \phi^{\alpha B}_{I}\,.
\end{align}
Since in this approach the supersymmetry transformation laws are described geometrically as Lie derivatives along the fermionic directions of superspace, the above procedure allows to easily determine the supersymmetry transformations of the fields, which read
\begin{align}\label{susytransf}\nonumber
  \delta_\varepsilon \phi^{\alpha A}_I \,=&\, \varepsilon^{(\alpha )A} \Lambda^\alpha_{(\alpha )I} \,, \\ \nonumber
  \delta_\varepsilon \Lambda^{\alpha (\alpha )}_I \,=& \,\Phi^{\alpha A}_{I;i}\left(\mathbb{M}^i_-\right)^{(\alpha)}{}_{(\beta)} \varepsilon^{(\beta)}_{A}+\mathbb N^\alpha_{IA} \varepsilon^{(\alpha)A} \,,\\ \nonumber
  \delta_\varepsilon e^i \,=&\,  \left(\mathbb{M}_-^i\right)_{(\alpha)(\beta)} \varepsilon^{(\alpha)A}\Psi^{(\beta)}_A  \,, \\ \nonumber
  \delta_\varepsilon\omega^i \,=&\, \frac{1}{L}(\mathbb M^i_+)_{(\alpha)(\beta)}\varepsilon^{(\alpha)A}\Psi^{(\beta)B)}\epsilon_{AB} \,, \\ \nonumber
 \delta_\varepsilon \Psi^{(\alpha)A}\,= & \nabla \varepsilon^{(\alpha)A}+\, \frac 1L\,\mathbb{K}_i^{(\alpha)(\beta)}\,e^i \varepsilon^A_{(\beta)}\,,\\
  \delta_\varepsilon A^x \,=&\, \frac \ii L(1+\alpha) (\mathbb{t}^x_{(3)})_{AB} \varepsilon^{(\alpha )A}\Psi_{(\alpha)}^B\,.
\end{align}
Note that the condition of background invariance under supersymmetry requires $\varepsilon$ to be a Killing spinor, namely that $\delta_\varepsilon {\Psi}^{(\alpha)A}=0$.  This in turn implies that the supersymmetry parameter should satisfy the following equation:
\begin{align}\label{KSE}
\hat{\nabla}\varepsilon \equiv \nabla \varepsilon +\, \frac 1L\,\mathbb{K}_i \,e^i\varepsilon =0\,. 
\end{align}
From this it follows that all the background fields have vanishing supersymmetry transformations on spacetime,
\begin{equation}\label{susyvarspacetime}
    \delta e^i_\mu= \delta A^x_\mu= \delta \omega^i_\mu = \delta \Psi^{(\alpha)A}_\mu =0\,.
\end{equation}

\section{The Lagrangian}
\label{lagrangian}
The geometric approach  allows to determine the Lagrangian for our dynamical hypermultiplets as a bosonic 3-form in the $\mathrm{D}^2 (2,1;\,\boldsymbol{\alpha})$ superspace. It reads
\begin{align}
   \mathcal{L}\,= & a_1 \left(\nabla \phi^{\alpha A}_{I} - \Psi^{(\alpha)A} \Lambda^\alpha_{(\alpha)I} \right) \, \Phi_{\alpha A}^{I;i} \, e^j \, e^k \epsilon_{ijk} - \frac 16 a_1 \Phi^{\alpha A}_{I;\ell} \Phi_{\alpha A}^{I;\ell} e^i \, e^j \, e^k \epsilon_{ijk} \nonumber\\
    &-16 \frac{ a_1}{\boldsymbol{\alpha}^2-1} \Lambda^{\alpha(\alpha)I} \left({\mathbb M}^i_+ \right)_{(\alpha) (\beta)} \biggl[ \frac 12 \nabla \Lambda^{\beta (\beta)}_I e^j + \left( \mathbb M^j_- \right)^{(\beta)}_{\phantom{(\beta)} (\gamma)} \left( \nabla \phi^{\beta A}_I  -\frac 12 \Psi^{(\delta) A} \Lambda^\beta_{(\delta) I} \right) \Psi^{(\gamma) B} \epsilon_{AB} \nonumber \\
    &- \mathbb N^\beta_{IA}(\phi) \Psi^{(\beta) A} e^j \biggr] \epsilon_{\alpha \beta} \, e^k \epsilon_{ijk} \nonumber \\
    &+ \frac{\ii a_1}{3(\boldsymbol{\alpha}+1)} \mathcal{M}_{(\alpha)(\beta)} \Lambda^{\alpha (\alpha) I} \Lambda^{\beta (\beta)}_I \epsilon_{\alpha \beta} e^i \,\, e^j \,\, e^k \epsilon_{ijk} -\frac{a_1}{3} \mathcal{V}(\phi)\,e^i\,\,e^j\,e^k \epsilon_{ijk} \nonumber\\
   &+\frac{\ii  a_1}{(\boldsymbol{\alpha}-1)} \phi^{\beta (A}_{I} \nabla \phi^{\alpha I |B)} \epsilon_{\alpha \beta}\,\Psi^{(\alpha) }_A \Psi^{(\beta) }_B\left[\frac{(1-\boldsymbol{\alpha}+\boldsymbol{\alpha}^2)}4  \delta_{(\alpha)(\beta)} +\boldsymbol{\alpha} \left( \mathbb T^k_{(1)} \mathbb T_{(2)k} \right)_{(\alpha)(\beta)} \right] \nonumber \\
    &- \frac{\ii (1+\boldsymbol{\alpha}) a_1}{8L} \phi^{IA \alpha} \phi^{\beta B}_I \left( \mathbb J^i \right)_{(\alpha) (\beta)} \Psi^{(\alpha)C} \Psi^{(\beta)D} e_i \epsilon_{CD} \epsilon_{\alpha \beta} \epsilon_{AB}\,.\label{hyperlag}
\end{align}
As we are going to show in  Section \ref{ward}, this Lagrangian is invariant under supersymmetry modulo boundary terms.\\
The hyperini mass matrix and the scalar potential have the following expressions:
\begin{align}
\mathcal{M}_{(\alpha)(\beta)}&=\frac 1L\left[\delta_{(\alpha)(\beta)}   +4\left(\mathbb{T}_{(1)}^i \cdot \mathbb{T}_{(2)i}\right)_{(\alpha)(\beta)} \right] \,, \label{mass}\\
\mathcal{V}(\phi)&={-} \frac{1}{2L^2} \phi^{\alpha A}_I\phi^{\beta B I} \epsilon_{\alpha \beta} \epsilon_{AB}+{\rm const.}\,.\label{pot}
\end{align}
We choose the overall normalization to be $a_1=\frac 12$. Thus, the spacetime projection of the superspace Lagrangian \eqref{hyperlag} takes the simple expression
\begin{align}
   \mathcal{L}_{\mbox{spacetime}}\,= & \frac 12 \nabla_\mu \phi^{\alpha A}_{I}\, \nabla^\mu \phi^{\beta BI} \epsilon_{\alpha\beta}\epsilon_{AB}\,-\frac{8}{\boldsymbol{\alpha}^2-1} \Lambda^{\alpha(\alpha)I} \left( {\mathbb M}_+^\mu \right)_{(\alpha) (\beta)}  \nabla_\mu \Lambda^{\beta (\beta)}_I\,\epsilon_{\alpha\beta}\nonumber \\
    &+ \frac{\ii}{(\boldsymbol{\alpha}+1)} \mathcal{M}_{(\alpha)(\beta)} \Lambda^{\alpha (\alpha) I} \Lambda^{\beta (\beta)}_I \epsilon_{\alpha \beta}  - \, \mathcal{V}(\phi)\,.\label{sptlag}
\end{align}
Notice that the spacetime Lagrangian describes non-mutually-interacting scalar and fermion sectors,  the interaction terms only appearing in the components of the superspace Lagrangian along the odd directions.\\
The expression of the scalar potential, which is in fact a mass term for the scalar fields, is fixed by the requirement of supersymmetry of the action, to be discussed in Section \ref{ward}. \par 
In this Section, we are going to explicitly write down the Euler-Lagrange equations of the spacetime Lagrangian, which provides the field equations of the hypermultiplets. Moreover, we are going to discuss some of the peculiarities of the superspace Lagrangian \eqref{hyperlag}, which are not apparent in its spacetime projection \eqref{sptlag}.

\subsection{The field equations}

The scalar field $\phi^{\alpha A}_I$ satisfies the following Klein-Gordon equation of motion: 
\begin{equation}
\nabla_\mu\nabla^\mu\,\phi^{\alpha A}_I={+} \frac 1{L^2}\phi^{\alpha A}_I \,,
\end{equation}
where the mass is given by the inverse of the AdS radius $L$.
Let us observe that the squared mass of the scalar fields, $m^2_\phi=-\frac 1{L^2}$, saturates the Breitenlohner-Freedman (BF) bound \cite{Breitenlohner:1982bm,Breitenlohner:1982jf} in $D=3$. Being the BF bound satisfied, the vacuum is perturbatively stable against scalar fluctuations.

The equation of motion of $\Lambda^{(\alpha) \alpha}_I$, that can be easily obtained from the Lagrangian \eqref{sptlag}, as well as (after some laborious calculations) from the Bianchi identities in superspace, reads
\begin{equation}
\label{eom_Lambda_new}
    \left( \mathbb{M}^i_{+} \right)_{(\alpha)(\beta)} \nabla_i \Lambda^{\alpha (\beta)}_I- \frac{\ii}{8} \left( \boldsymbol{\alpha}-1 \right) \mathcal{M}_{(\alpha)(\beta)}  \Lambda^{\alpha (\beta)}_I= 0 \,.
\end{equation}
It is a massive Dirac  equation, with a constant mass proportional to the inverse of the AdS radius $L$. The mass matrix $\mathcal M$ can be diagonalized
through conjugation with the orthogonal matrix
\begin{equation}
    \mathcal{P}=\left( \begin{array}{cccc}
        0 & 0 & 0 & 1 \\
        -\frac{1}{\sqrt{2}} & 0 & \frac{1}{\sqrt{2}} & 0 \\
        \frac{1}{\sqrt{2}} & 0 & \frac{1}{\sqrt{2}} & 0 \\
        0 & 1 & 0 & 0
    \end{array} \right) \,,
\end{equation}
showing that it has only one eigenvalue different from zero,
\begin{equation}\label{D}
    \mathcal{M_D}_{(\sigma)(\alpha)}=\left( \mathcal{P}^t \mathcal{M} \mathcal{P} \right)_{(\sigma)(\alpha)}= \frac 1L\,\left( \begin{array}{cccc}
        4 & 0 & 0 & 0 \\
        0 & 0 & 0 & 0 \\
        0 & 0 & 0 & 0 \\
        0 & 0 & 0 & 0 \\
    \end{array} \right) \,. \\
\end{equation}
The interpretation of the above result in terms of mass eigenstates will be more transparent in the  twisted descriptions of the model that we will give in Section \ref{twist}.

\subsection{Scalar potential and supersymmetry invariance}\label{ward}
In this Section, we discuss the supersymmetry of the action of our model, starting from the properties of the Lagrangian both in superspace \eqref{hyperlag} and in spacetime \eqref{sptlag}.
As we are going to see, we find that supersymmetry invariance of the superspace Lagrangian requires a non-trivial contribution from the boundary. This means that the bulk Lagrangian is invariant modulo total derivative terms. The latter are relevant to the complete invariance of the action, being our model formulated on a spacetime with AdS geometry, which is not globally hyperbolic. Here, we will be dealing with the invariance of the model in the bulk only, leaving a detailed analysis of the invariance of the action, which includes the boundary contributions, along the lines of \cite{Andrianopoli:2014aqa}, to future investigation.
For this reason, we expect all contributions $\mathcal Y$ in $\delta\mathcal L$ to sum up to a total derivative term $\diff\delta\mathcal{Z}$ in such a way that
\begin{align}
    \delta\mathcal L=\mathcal Y=\underbrace{\mathcal Y+ \diff (\delta\mathcal Z)}_0-\diff (\delta\mathcal Z)=-\diff (\delta\mathcal Z)\,.\label{quasisymmetry}
\end{align}
By using the transformation laws \eqref{susytransf}, restricted to spacetime and the Killing spinor equation \eqref{KSE}, it can be verified that  the spacetime Lagrangian \eqref{sptlag} features off-shell invariance under supersymmetry. \\
In particular, this invariance is crucial to determine the explicit expression of the scalar potential appearing in \eqref{hyperlag} and \eqref{sptlag}, as expected. Indeed, invariance of the spacetime Lagrangian \eqref{sptlag} to order $1/L^2$ requires the scalar potential $\mathcal{V}(\phi)$ to satisfy the following condition:
\begin{equation}
    \frac{\partial \mathcal{V}}{\partial \phi^{\alpha A}}=-\frac{1}{L^2}\,\epsilon_{\alpha\beta}\epsilon_{AB}\,\phi^{\beta B}\,,
\end{equation}
which yields the expression for the potential given in \eqref{pot}.

The analysis of supersymmetry for the superspace Lagrangian can instead be performed, in a geometric setting, by computing its Lie derivative along odd diffeomorphisms,
\begin{equation}
\delta_\epsilon \mathcal{L}= \pounds_\epsilon \mathcal{L}  =  \iota_\epsilon (\diff \mathcal{L})+\diff \iota_\epsilon (\mathcal{L}) \,,
\end{equation}
and ignoring the total derivative part for the bulk analysis, as explained above. 
Eventually, we can analyse independently the invariance in  different sectors, defined by the inverse powers of the AdS radius $L$ and on different basis elements for 3-forms in superspace.
Of particular interest is the sector $\frac{1}{L^2}\epsilon \Psi ee $, which  yields the supersymmetric potential Ward identity \cite{Ferrara:1985gj,Cecotti:1985mx,Andrianopoli:2015rpa}. \\
The explicit computation of this sector in \eqref{hyperlag} yields
\begin{align}\nonumber
    &\mathcal V(\phi)\left((\boldsymbol\alpha+1)\mathbb K^i-(\boldsymbol\alpha-1)\mathbb J^i\right)+\frac{1}{2L^2}\frac{\boldsymbol{\alpha}+1}{\boldsymbol{\alpha} -1}\phi^2\left((1-\boldsymbol\alpha)\mathbb K^i+(\boldsymbol\alpha+1)\mathbb J^i\right)\\
    &-\frac{\boldsymbol\alpha+1}{L^2}\phi^2\epsilon^{ijk}\mathbb K_j\mathbb J_k=\left.\mathcal{Y}\right\vert_{\frac{1}{L^2}\epsilon \Psi  e e }\neq 0\,.
\end{align}
In particular, we notice that, while the components on the left hand side along $\mathbb{K}^i$ vanish for the choice of the potential in \eqref{pot}, the components along $\mathbb{J}^i$ fail to do so.
These contributions can be disposed of by adding a suitable total derivative term to the superspace Lagrangian in \eqref{hyperlag} of the form
\begin{align}\label{totalderivative} \diff\mathcal Z=\diff \left( \Lambda^{\alpha(\alpha)I} \left[r_1\left( \mathbb J^i \right) + r_2\epsilon^{ijk}\mathbb T_{(1)j} \mathbb T_{(2)k} \right]_{(\alpha) (\beta)} \phi^{\beta A}_{I} \Psi^{(\beta)B} \epsilon_{AB} \epsilon_{\alpha \beta} e_i \right)\,,
\end{align}
where the values of $r_1,r_2$ are restricted by  
the requirement that the $\frac{1}{L^2}\epsilon \Psi ee $ component of $\delta_\epsilon \mathcal{L}$ vanishes,
\begin{align}\nonumber
    &\mathcal V(\phi)\left((\boldsymbol\alpha+1)\mathbb K^i-(\boldsymbol\alpha-1)\mathbb J^i\right)+\frac{1}{2L^2}\frac{\boldsymbol{\alpha}+1}{\boldsymbol{\alpha} -1}\phi^2\left((1-\boldsymbol\alpha)\mathbb K^i+(\boldsymbol\alpha+1)\mathbb J^i\right)\\
    &-\frac{\boldsymbol\alpha+1}{L^2}\phi^2\epsilon^{ijk}\mathbb K_j\mathbb J_k+ \frac{2(\boldsymbol\alpha+1) }{ L^2}\left(r_1+\frac 12 r_2\right)\phi^2\mathbb J^i=0\,.
\end{align}
This leads to the following relation:
\begin{equation}
 r_1+\frac{r_2}{2}=-\frac{1}{2} \left(\frac{\boldsymbol\alpha^2+1}{\boldsymbol\alpha^2-1}\right) \,.
\end{equation}
In light of the remark in \eqref{quasisymmetry}, this signals that the Lagrangian in \eqref{hyperlag} is invariant in the bulk modulo a total derivative term $-\diff (\delta \mathcal{Z})$.

\subsection{Comments on the dependence of the Lagrangian on the hyper-K\"ahler geometry}\label{hyperkahlerstructure}

Although we are restricting to a flat hyper-K\"ahler manifold, in view of a possible generalisation to a curved one, it would be useful to provide an intrinsic characterization of the scalar dependence of the Lagragian (in particular of the scalar potential) in terms of quantities characterizing the hyper-K\"ahler geometry. To this end we start recalling the main facts about hyper-K\"ahler geometry.\par
A hyper-K\"ahler manifold \cite{Hitchin:1986ea,Galicki:1986ja} of real dimension $4n_H$ is a manifold on which three complex structures are defined $J^x$, $x=1,2,3$, $(J^x)^2=-{\bf 1}$, closing an $\mathfrak{su}(2)$ algebra: $[J^x,\,J^y]=\epsilon^{xyz}\,J^z$. The metric $h_{uw}$ is required to be \emph{Hermitian} with respect to any of the three structures. In a local patch with coordinates $q^u$, $u=1,\dots, 4n_H$, this amounts to the conditions
\begin{equation}
  h_{uw}\,J^{x\,w}{}_v+h_{vw}\,J^{x\,w}{}_u=0\,,
\end{equation}
where $(J^x)^u{}_v$ represent the action of the complex structures on a coordinate basis of the tangent space and satisfy the quaternionic algebra
\begin{equation}
    J^xJ^y=-\delta^{xy}+\epsilon^{xyz}\,J^z\,.
\end{equation}
The manifold is further required to be K\"ahler with respect to each of the three complex structures. This in turn is equivalent to the condition that
 the matrices $(J^x)^u{}_v$ be covariantly constant with respect to the Levi-Civita connection $\tilde{\Gamma}^w_{uv}$ on the manifold,
\begin{equation}
    \mathcal{D}_w\,(J^x)^u{}_v=0\,.
\end{equation}
We define the hyper-K\"ahler 2-forms as follows:
\begin{equation}
    \Omega^x=\Omega^x_{uv}\,\diff q^u\wedge \diff q^v\,\,,\,\,\,\,\Omega^x_{uv}=h_{uw}\,J^{x\,w}{}_v=h_{w[u}\,J^{x\,w}{}_{v]}\,.
\end{equation}
The hyper-K\"ahler condition implies that these three 2-forms are closed: $\diff \Omega^x=0$.\par
In our case the coordinates $q^u$ are identified with the scalar fields of our model $\phi^{\alpha A}_I$.
The matrices $(J^x)^u{}_v$ are constant and define the (linear) action of the $\mathfrak{su}(2)$ generators on the index $A$ of the scalars  $\phi^{\alpha\,A}_I$,
\begin{equation}
(J^x)^u{}_v=(J^x)^{A\alpha I}{}_{B\beta J}=2\,(\mathbb{t}^x)^A{}_B\,\delta^\alpha_\beta\,\delta^I_J\,.
\end{equation}
The real dimension of the space is $8$, corresponding to $n_H=2$ hypermultiplets.\par
We can treat the space as a complex manifold with respect to the complex structure $J\equiv J^{x=2}$, which acts on the indices $A,\,B,\,\dots$ as the matrix $2\,(\mathbb{t}^{{2}})^A{}_B=\ii \,(\sigma^2)^A{}_B$ 
\begin{equation}
J\cdot \phi^{\alpha A}_I={\ii \,(\sigma^2)^A{}_B}\,\phi^{\alpha B}_I\,.
 \end{equation}
This choice of the complex structure yields the definition of 4 complex coordinates $\phi^{\alpha}_I$ and their complex conjugates $\bar{\phi}^{\alpha}_I$,
\begin{equation}
\phi^{\alpha}_I\equiv \phi^{\alpha A=1}_I+\ii\,\phi^{\alpha A=2}_I\,,\,\,\,\,\,\bar{\phi}^{\alpha}_I\equiv\phi^{\alpha A=1}_I-\ii\,\phi^{\alpha A=2}_I\,,
\end{equation}
 such that
\begin{equation}
 J\cdot \diff \phi^{\alpha}_I=-\ii\,\diff \phi^{\alpha}_I\,\,,\,\,\,J\cdot \diff \bar{\phi}^{\alpha}_I=\ii\,\diff \bar{\phi}^{\alpha}_I\,.
\end{equation}
When interpreting $\phi^{\alpha}_I,\,\bar{\phi}^{\alpha}_I$ as ghost and anti-ghost fields, the operator $\ii\,J$ measures their \emph{ghost charges}, which are $+1$ and $-1$, respectively. In our model the hyper-K\"ahler manifold is flat and the metric reads
\begin{equation}
ds^2=h_{uv}\,\diff q^u\,\diff q^v=\epsilon_{\alpha\beta}\,\epsilon_{AB}\,\diff \phi^{\alpha A}_I\,\diff {\phi}^{\beta B}_I=\ii\epsilon_{\alpha\beta}\,\diff \phi^\alpha_I\,\diff \bar{\phi}^\beta_I \,.
\end{equation}
The K\"ahler 2-form associated with $J$ reads
\begin{equation}
K=h_{uv}\,J^{v}{}_w\,\diff q^u\wedge \diff q^w=-\diff \phi^{\alpha}\wedge \diff \bar{\phi}^\beta\,\epsilon_{\alpha\beta}\,,
\end{equation}
and the corresponding K\"ahler potential has the following expression:
\begin{equation}\label{kahler}
 \mathcal {K}(\phi,\bar{\phi})\equiv \phi^{\alpha A}_I\phi^{\beta B}_I \epsilon_{\alpha \beta} \epsilon_{AB}=\ii\,\epsilon_{\alpha\beta}\,\phi^\alpha_I\,\bar{\phi}^\beta_I\,.
\end{equation}
In terms of this potential the metric in the complex basis is given by the known relation for K\"ahler manifolds,
\begin{equation}
\diff s^2=\left(\frac{\partial^2 \mathcal {K}}{\partial\phi^\alpha_I\partial \bar{\phi}^\beta_J}\right)\,\diff\phi^\alpha_I\,\diff\bar{\phi}^\beta_J\,.
\end{equation} 
As for the other complex structures $J^x$, whose action on the $A,\,B$ indices can be described in terms of the matrices $2\,(\mathbb{t}^x)^A{}_B$, it is useful to describe the index $x=1,2,3$ in terms of a symmetric couple $(AB)$ and write $(J^x)^C{}_D=\ii\,(\mathbb{t}^x)_{AB}\,(J^{(AB)})^C{}_D$, where $(J^{(AB)})^C{}_D\equiv \delta^{(A}_D\epsilon^{B)C}$. The three closed hyper-K\"ahler 2-forms  $\Omega^{(AB)}$ have then the following expression \cite{Andrianopoli:2019sqe}:
\begin{equation}
    \Omega^{(AB)}=\diff\phi^{\alpha\,A}_I\wedge \diff \phi^{\beta\,B}_I\epsilon_{\alpha\beta}\,.
\end{equation}
Being closed, locally these forms can be written as the exterior derivative of 1-forms $\mathcal{A}^{(AB)}$: $ \Omega^{(AB)}=\diff \mathcal{A}^{(AB)}$, where
\begin{equation}
\mathcal{A}^{(AB)}=\phi^{\alpha\,(A}_I\,\diff \phi^{\beta\,|B)}_I\epsilon_{\alpha\beta}\,.
\end{equation}
Let us now show that the dependence of the Lagrangian on the scalar fields can be described in terms of geometrical quantities which are intrinsic to the hyper-K\"ahler manifold and this suggests a natural generalisation of its expression to more general non-flat hyper-K\"ahler geometries \cite{Rozansky:1996bq}.
We note indeed that the scalar potential $\mathcal{V}(\phi,\bar{\phi})$ can be expressed in terms of $\mathcal{K}(\phi,\bar{\phi})$ as follows:
\begin{equation}
\mathcal{V}(\phi,\bar{\phi})=-\frac{1}{2L^2}\,\mathcal{K}(\phi,\bar{\phi})+{\rm const.}\,.
\end{equation}
Moreover the expression $\phi^{\alpha\,(A}_I\,\nabla \phi^{\beta\,|B)}_I\epsilon_{\alpha\beta}$ in a $\Psi\Psi$-component of the Lagrangian, as well as $\nabla \phi^{\alpha A}_I$ in the spacetime Lagrangian, are respectively interpreted in terms of the connection  $\mathcal{A}^{(AB)}$ and the vielbein $\mathcal{U}^{\alpha A}_I=\mathcal{U}^{\alpha A}_{I\,u}\,\diff q^u$ 1-forms, in which the exterior derivative $\diff$ is replaced by the covariant one $\nabla$ due to the gauging of the ${\rm SU}(2)$ isometry algebra by $A^x_\mu$.
As mentioned above, this observation is useful in view of a generalisation of the Lagrangian to a sigma-model on more general hyper-K\"ahler manifolds \cite{Rozansky:1996bq}. This task will be undertaken in a future work.

Let us add that, when a curved hyper-K\"ahler manifold is considered, the supersymmetry transformation laws contain extra contributions depending on the affine connection on the $\sigma$-model and the Lagrangian includes an additional term of the form
\begin{equation}
\epsilon^{AB}\,R_{\alpha A I,\beta B J; \gamma K, \sigma L}\,\Lambda^{(\alpha) \alpha I}\,\Lambda^{(\beta) \beta J}\,\Lambda^{(\gamma) \gamma K}\,\Lambda^{(\sigma) \sigma L}\epsilon_{(\alpha)(\beta)(\gamma)(\sigma)}\,,
\end{equation}
where
\begin{equation}
\epsilon_{(\alpha)(\beta)(\gamma)(\delta)}=4\,(\mathbb{T}_{(1)}^i)_{[(\alpha)(\beta)}(\mathbb{T}_{(1)\,i})_{(\gamma)(\delta)]}=-4 (\mathbb{T}_{(2)}^i)_{[(\alpha)(\beta)}(\mathbb{T}_{(2)\,i})_{(\gamma)(\delta)]}
\end{equation}
is the totally antisymmetric ${\rm SO}(2,2)$-invariant tensor and $$R_{\gamma K, \sigma L}=\frac{1}{2}\,R_{uv;\gamma K, \sigma L}\diff q^u\wedge \diff q^v= \frac{1}{2}\,R_{\alpha A I,\beta B J; \gamma K, \sigma L}\,\diff \phi^{\alpha A I} \wedge \diff \phi^{\beta BJ}$$ is the curvature 2-form with value in the $\mathfrak{usp}(2n)=\mathfrak{usp}(4)$ algebra.

\section{The Twists}
\label{twist}
In this Section we will perform two different twists of the theory. As we shall see, the first one will relate this model to the one of \cite{Rozansky:1996bq}, whereas the second one will allow to make contact with the unconventional supersymmetry explored in \cite{Alvarez:2011gd,Andrianopoli:2018ymh,Andrianopoli:2019sip,Andrianopoli:2020zbl,Andrianopoli:2019sqe}.

\subsection{{First twist}}

In all of the analysis up to now, the manifest invariance of the action is only restricted to the Lorentz group ${\rm SL}(2,\mathbb{R})_L$ embedded as the diagonal subgroup
$${\rm SL}(2,\mathbb{R})_L={\rm SL}(2,\mathbb{R})'_D\subset {\rm SL}(2,\mathbb{R})'_1\times{\rm SL}(2,\mathbb{R})'_2 \,$$ and to the R-symmetry group ${\rm SU}(2)$ inside $ \mathrm{D}^2 (2,1;\,\boldsymbol{\alpha})$. However, we have kept so far a somewhat hybrid notation in the description of the fermionic fields and of supersymmetry, by keeping the spinor indices $\alpha',\dot{\alpha}'$ of ${\rm SL}(2,\mathbb{R})'_1$ and ${\rm SL}(2,\mathbb{R})'_2$ distinct and thus working with a redundant 4-component description of spinor fields.\\
In this Subsection we rewrite the spinor fields in irreducible ${\rm SL}(2,\mathbb{R})'_D$ components,
\begin{align}
\Lambda^{(\alpha)\alpha}_I\,&\rightarrow\,\,\varPsi_{I i}^\alpha\,,\,\,\,\eta^\alpha_I\,,
\end{align}
according to the branching
\begin{align}
\left({\bf 2},\,{\bf 2}\right)\,&\rightarrow \,\,{\bf 3}+{\bf 1}\,.
\end{align}
We will call this decomposition a \emph{twist} in analogy with the known \emph{topological twist}. In our framework, it amounts to making explicit the choice of the spin-connection of  $\mathrm{D}^2(2,1;\,\boldsymbol{\alpha})$ superspace among the  $\rm{SL}(2,\mathbb{R})$ connections of the superalgebra. 
\par
To express $\Lambda^{(\alpha)\,\alpha}_I$ in terms of its component fields, we introduce the following intertwining matrices:
\begin{equation}
\gamma^i_{(\alpha)}\equiv (\gamma^i)_{\alpha'\dot{\alpha}'},\,\,\,\epsilon_{(\alpha)}\equiv \epsilon_{\alpha'\dot{\alpha}'}\,,
\end{equation}
which are clearly invariant under the Lorentz group ${\rm SL}(2,\mathbb{R})_L$.\footnote{To see this for $\gamma^i_{(\alpha)}$ one can verify that
$$(\mathbb{T}^i_{(1)}+\mathbb{T}^i_{(2)})^{(\alpha)}{}_{(\beta)}\,\gamma^{k\,(\beta)}=-
\epsilon^{ik\ell}\,\gamma_\ell^{(\alpha)}\,.$$}
Using this quantity, we can decompose $\Lambda^{\alpha (\alpha)}_I$ into the components  $\varPsi^{\alpha}_{i\,I},\,\eta^\alpha_I$ as follows:
\begin{equation}
    \Lambda^{\alpha (\alpha)}_I= \ii\,\gamma^{i(\alpha)}\,\varPsi^{\alpha}_{i\,I}+\epsilon^{(\alpha)}\,\eta^\alpha_I\,.\label{twist1}
\end{equation}

In the context of the analysis carried out in \cite{Andrianopoli:2019sqe}, the two components of $ \Lambda^{\alpha (\alpha)}_I$ resulting from the twist describe, respectively, the gauge field $\varPsi^{\alpha}_{i\,I}$ associated with the odd gauge symmetries of a Chern-Simons model defined on the supergroup ${\rm OSp}(2|2)$ and the corresponding Nakanishi-Lautrup field $\eta^\alpha_I$.
In our case, similarly as in \cite{Rozansky:1996bq}, the bosonic subgroup of the gauge supergroup being replaced by a global flavour symmetry, the odd gauge fields $\varPsi^{\alpha}_{i\,I}$ are the only relics of the Chern-Simons gauge supergroup. Similarly, the analogous components of the supersymmetry generators $\mathcal{Q}^{(\alpha)A}$ define what in \cite{Kapustin:2009cd} were identified as the BRST symmetry generator $\mathcal{S}$, the ``vector'' BRST symmetry generator $\mathcal{S}_i$ and their secondary counterparts $\bar{\mathcal{S}},\,\bar{\mathcal{S}}_i$:
\begin{equation}
   \mathcal{Q}^{ (\alpha)A}= \ii\,\gamma^{i(\alpha)}\,\mathcal{S}^{A}_{i}+\epsilon^{(\alpha)}\,\mathcal{S}^A\,,
\end{equation}
where $\mathcal{S}^A\equiv (\mathcal{S},\bar{\mathcal{S}}),\,\mathcal{S}^A_i\equiv (\mathcal{S}_i,\bar{\mathcal{S}}_i)$.

Let us now compute the anticommutator of the two supersymmetry generators for the $\mathrm{D}^2(2,1;\,\boldsymbol{\alpha})$ algebra in \eqref{algebra}, in terms of  the twisted operators $\mathcal{S}^A$, $\mathcal{S}^A_i$. We find
\begin{align}
\left\{\mathcal{S}^A,\mathcal{S}^B\right\}&= \frac{\ii}{2}\,s_3\,\left(\mathbb{t}_{(3)x}\right)^{AB}\,\mathcal{T}_{(3)}^x \,, \quad
\label{algebratwist1}\\
\left\{\mathcal{S}^A_i,\mathcal{S}^B_j\right\}&=\frac \ii 4 \epsilon_{ijk}\epsilon^{AB}\left(s_1 \mathcal{T}_{(1)}^k+s_2 \mathcal{T}_{(2)}^k\right) + \frac \ii 2\,s_3\,\eta_{ij}\,\left(\mathbb{t}_{(3)x}\right)^{AB}\,\mathcal{T}_{(3)}^x \,, \quad 
\label{algebratwist2}\\
\left\{\mathcal{S}^A_i,\mathcal{S}^B\right\}&=\frac \ii {4} \epsilon^{AB}\left(s_1 \mathcal{T}_{(1)i} -s_2 \mathcal{T}_{(2)i}\right)\,
\,.\label{algebratwist3}
\end{align}
The above expressions show that, except for the $\mathrm{D}^2(2,1;\,\boldsymbol{\alpha})$ singular value $s_3= 0$ (corresponding to $\boldsymbol{\alpha}=-1$), the scalar generators $\mathcal{S}^A$ do not behave as cohomology operators.

By proving the following relation:
\begin{equation}
    (\mathbb{T}^i_{(1)}\mathbb{T}_{(2)\,i})^{(\alpha)}{}_{(\beta)}\,\gamma^{k\,(\beta)}=-\frac{1}{4}\,\gamma^{k\,(\alpha)}\,,
\end{equation}
one can verify that $\gamma^{i\,(\alpha)}$ provide three zero-eigenvectors for the mass matrix $\mathcal{M}_{(\alpha)(\beta)}$:
\begin{equation}
    \mathcal{M}^{(\alpha)}{}_{(\beta)}\gamma^{i(\beta)}=\left(\delta^{(\alpha)}_{(\beta)}+4\,(\mathbb{T}^j_{(1)}\mathbb{T}_{(2)\,j})^{(\alpha)}{}_{(\beta)}\right)\gamma^{i(\beta)}=0\,.
\end{equation}
This implies that the massive degrees of freedom are encoded in $\eta^\alpha_I$.
We should now write the equation for $\Lambda$ in terms of $\varPsi$ and $\eta$.
To this end it is useful to write the following relations (we suppress the indices $\alpha,\,I$):
\begin{align}
    (\mathbb{T}^i_{(1)}\Lambda)^{(\alpha)}&=\frac{{\rm i}}{2}\,\epsilon^{i\ell k}\,\gamma^{(\alpha)}_\ell\,\varPsi_k-\frac{1}{2} \epsilon^{(\alpha)}\varPsi^i+\frac{{\rm i}}{2}\,\gamma^{i(\alpha)}\,\eta\,,\nonumber\\
   (\mathbb{T}^i_{(2)}\Lambda)^{(\alpha)}&=\frac{{\rm i}}{2}\,\epsilon^{i\ell k}\,\gamma^{(\alpha)}_\ell\,\varPsi_k+\frac{1}{2} \epsilon^{(\alpha)}\varPsi^i-\frac{{\rm i}}{2}\,\gamma^{i(\alpha)}\,\eta\,.
\end{align}
By substituting \eqref{twist1} in the field equation \eqref{eom_Lambda_new} and projecting along $\left( \gamma_p \right)^{(\sigma)}$ and $\epsilon^{(\sigma)}$ we find\footnote{We use the following identity:
\begin{align}
    \left( \gamma_p \right)^{(\sigma)} \epsilon_{(\sigma)} = 0 , \quad \left( \gamma_p \right)^{(\sigma)} \left( \gamma_k \right)_{(\sigma)} = -2\eta_{pk} , \quad \epsilon^{(\sigma)} \epsilon_{(\sigma)} = 2 \,. \nonumber
\end{align}}
\begin{align}
    \left( \gamma_p \right)^{(\sigma)} &: \quad \left( \boldsymbol{\alpha} -1\right) \nabla_p \eta_I +  \left( 1+ \boldsymbol{\alpha} \right) \epsilon_{pik} \nabla^i \varPsi_I^{k} = 0\,, \nonumber\\
    \epsilon^{(\sigma)} &: \quad \left( \boldsymbol{\alpha} - 1 \right) \nabla_i \varPsi_I^{i } + \frac{2}{L} \left( \boldsymbol{\alpha} - 1 \right) \eta_I = 0 \,.\label{eqpsieta}
\end{align}
We observe that for the value $\boldsymbol{\alpha}=1${, which is not a singular case for $\mathrm{D}^2 (2,1;\,\boldsymbol{\alpha})$},\footnote{In fact, for $\boldsymbol{\alpha}=1$ the algebra $\mathrm{D}^2 (2,1;\,\boldsymbol{\alpha})$ becomes isomorphic to a real form of $\mathrm{D}(2,1)\sim \mathfrak{osp}(4|2)$ with bosonic subgroup $\mathfrak{so}(2,2)\times \mathfrak{su}(2)$.} $\eta$ decouples and we end up with only one equation for $\varPsi^\alpha_i$. Note that if the index $\alpha$ were a spinor index with respect to the Lorentz group, the equation $\epsilon_{ipk} \nabla^p \varPsi_I^{k \alpha} = 0$ would be the Rarita-Schwinger equation for a massless spin-$3/2$ field. Recall however that in our construction $\alpha$ is an internal gauge index.
{\paragraph{The Lagrangian and the supersymmetry variations\\}
The above mentioned equations of motion can be reproduced by the following Lagrangian: 
\begin{align}\nonumber
\mathcal L_{\rm spacetime}=& { \frac 12 \nabla_{{i}} \phi^{\alpha A}_{I}\, \nabla^{{i}} \phi^{\beta B I} \epsilon_{\alpha\beta}\epsilon_{AB}-\mathcal V(\phi)+\frac{4 \ii}{1-\boldsymbol{\alpha}}\epsilon^{ijk}\varPsi^{\alpha I}_i\nabla_j\varPsi^\beta_{kI}\epsilon_{\alpha\beta} } \nonumber \\
&{+\frac {8 \ii}{(1+\boldsymbol{\alpha})}\left(-\varPsi^{i\alpha I}\nabla_i \eta^\beta_I
+\frac{1}{L}\eta^{\alpha I}\eta^\beta_I\right)\epsilon_{\alpha\beta}\,,}
\end{align}
which can be obtained from the spacetime Lagrangian \eqref{sptlag} by performing the twist \eqref{twist1}. Note that the $\eta_I$-dependent terms in the above Lagrangian are consistent with the interpretation of $\eta_I$ as the Nakanishi-Lautrup field \cite{Andrianopoli:2019sqe}.\\
The supersymmetry variations of these two new fields and of $\phi$ are
\begin{align}
\delta_\varepsilon \phi^{\alpha A}_I &= \varepsilon^{(\alpha) A} \left( \ii \gamma^i_{(\alpha)} \varPsi^\alpha_{iI} + \epsilon_{(\alpha)} \eta^\alpha_I \right) \,, \nonumber \\
\delta_\varepsilon \varPsi^\alpha_{i I}&=-\frac18\left((1-\boldsymbol{\alpha})\epsilon_{ijk}\nabla^j\phi^{\alpha A}_I+\frac{1+\boldsymbol{\alpha}}{L}\phi^{\alpha A}_I\eta_{ik}\right)\gamma^k_{(\alpha)}\varepsilon^{(\alpha)}_A-\frac{\ii(1+\boldsymbol{\alpha})}{8}\nabla_i\phi^{\alpha A}_I\epsilon_{(\alpha)}\varepsilon^{(\alpha)}_A \,, \nonumber \\
\delta_\varepsilon \eta^\alpha_{ I}&=-\frac{1+\boldsymbol{\alpha}}{8}\nabla_i\phi^{\alpha A}_I\gamma^i_{(\alpha)}\varepsilon^{(\alpha)}_A+ {\frac{1}{2}} \mathbb N^\alpha_{IA}\epsilon_{(\alpha)}\varepsilon^{(\alpha)A} \,.
\end{align}
These expressions can now be rewritten in terms of the new symmetry parameters arising from the twist we are considering, $\varepsilon^{(\alpha)A}=\ii\gamma^{i(\alpha)}\varepsilon^A_i+\epsilon^{(\alpha)}\varepsilon^A$, that is 
\begin{align}\label{variationssplit}
\delta_{\varepsilon^{iB}\mathcal S_{iB}}\phi^{\alpha A}_I&=\varepsilon^{iA}\varPsi^\alpha_{iI} \,, \nonumber \\
\delta_{\varepsilon^{B}\mathcal S_{B}}\phi^{\alpha A}_I&=\epsilon^{AB}\eta^{\alpha}_I\varepsilon_B \,, \nonumber \\
\delta_{\varepsilon^{lB}\mathcal S_{lB}}\varPsi^{\alpha}_{iI}&=\frac{\ii}{8}\left((1-\boldsymbol{\alpha})\epsilon_{ijk}\nabla^j\phi^{\alpha A}_I+\frac{1+\boldsymbol{\alpha}}{L}\phi^{\alpha A}_I\eta_{ik}\right)\varepsilon^k_A \,, \nonumber\\
\delta_{\varepsilon^{B}\mathcal S_{B}}\varPsi^{\alpha}_{iI}&=-\frac{\ii(1+\boldsymbol{\alpha})}{8}\nabla_i\phi^{\alpha A}_I\varepsilon_A \,, \nonumber \\
\delta_{\varepsilon^{iB}\mathcal S_{iB}} \eta^{\alpha}_{I}&=\frac{\ii(1+\boldsymbol{\alpha})}{8}\nabla_i\phi^{\alpha A}_I\varepsilon_A^i \,, \nonumber \\
\delta_{\varepsilon^{B} \mathcal S_{B}}\eta^{\alpha}_{I}&=\frac{\ii(1+\boldsymbol{\alpha})}{8 {L}}\phi^{\alpha}_{AI}\varepsilon^A \,.
\end{align}
In particular, since on a generic field $\varPhi$ we have by linearity $\delta_{\varepsilon^{B}\mathcal S_{B}}\varPhi=\varepsilon^{B}\delta_{\mathcal S_{B}}\varPhi\equiv \varepsilon^{B}(\mathcal{S}_B\cdot \varPhi)$, from the above equations we obtain
\begin{align}\label{operatorS}
(\mathcal{S}^B\cdot \phi^{\alpha A}_I)&=\epsilon^{AB}\eta^{\alpha}_I \,, \nonumber\\
(\mathcal{S}^A\cdot \varPsi^{\alpha}_{iI})&= \frac{\ii(1+\boldsymbol{\alpha})}{8}\nabla_i\phi^{\alpha A}_I \,, \nonumber\\
(\mathcal{S}^A\cdot \eta^{\alpha}_I)&=\frac{\ii(1+\boldsymbol{\alpha})}{8{L}}\phi^{\alpha A}_I \,,
\end{align}
so that
\begin{align}\label{operatorSS}
(\mathcal{S}^{(B}\mathcal{S}^{A)}\cdot \phi^{\alpha C}_I)&=\frac{\ii(1+\boldsymbol{\alpha})}{8 L}\epsilon^{C(A}\phi^{\alpha B)}_I \,, \nonumber\\
(\mathcal{S}^{(B}\mathcal{S}^{A)}\cdot \varPsi^{\alpha}_{iI})&= \frac{\ii(1+\boldsymbol{\alpha})}{8}\epsilon^{(AB)}\nabla_i\eta^{\alpha }_I=0 \,, \nonumber\\
(\mathcal{S}^{(B}\mathcal{S}^{A)}\cdot \eta^{\alpha}_I)&=\frac{\ii(1+\boldsymbol{\alpha})}{8 {L}}\epsilon^{(AB)}\eta^{\alpha }_I=0 \,.
\end{align}
The above equations show that the operators $\mathcal{S}^A$ do not anticommute on fields with non-vanishing ghost-number. The cohomological structure is retrieved in the singular case $\boldsymbol{\alpha}+1=0$. \\
Furthermore, we notice that the obtained spacetime Lagrangian can be expressed, as in \cite{Rozansky:1996bq} 
\begin{align}
\mathcal L_{\rm spacetime}=&\frac{4 \ii}{1+\boldsymbol{\alpha}} \mathcal{S}^A\cdot \left( \nabla^i \phi^{\alpha B I} \varPsi^\beta_{i I} \epsilon_{AB} + \frac{1}{L} \eta^{\alpha I} \phi^{\beta B}_I \epsilon_{AB} \right) \epsilon_{\alpha \beta} +\frac{4 \ii}{1-\boldsymbol{\alpha}}\epsilon^{ijk}\varPsi^{\alpha I}_i\nabla_j\varPsi_{k\alpha I} \,.
\end{align}
{Let us remark that the charges $\mathcal{S}^A$ act on the spacetime Lagrangian similarly as BRST cohomology operators, separating it in a ``physical'' Lagrangian and a term in the image of $\mathcal{S}$, 
despite the fact that the $\mathcal{S}^A$ do not behave as proper cohomological charges (see \eqref{algebratwist1}). In fact, extra contributions due to the peculiar structure of the $\mathrm{D}^2(2,1;\,\boldsymbol{\alpha})$ superspace show up in the superspace Lagrangian, some of them being associated with the commutator $[\mathcal{S}^A,\nabla]\phi^{B\beta}_I$, which however vanishes on spacetime, as a consequence of \eqref{susyvarspacetime}, whose twisted expression implies a trivial action on spacetime of the $\mathcal{S}^A$ on the background fields appearing in the covariant derivatives.
}}

\subsection{{Second twist}}

As discussed in \cite{Andrianopoli:2019sqe}, in order to make contact with the model of \cite{Alvarez:2011gd}, where an ``unconventional'' supersymmetric theory featuring spin-$1/2$ fields $\chi_I^{(\rm{AVZ})}$ was constructed, we perform a second twist which amounts to writing the fields in a covariant way with respect to the diagonal subgroup ${\rm SL}(2,\mathbb{R})_D$ of ${\rm SL}(2,\mathbb{R})\times {\rm SL}(2,\mathbb{R})'_1$, where the former factor is the flavor symmetry group acting on the index $\alpha$. This amounts to introducing the ${\rm SL}(2,\mathbb{R})_D$-invariant tensors $\gamma^i_{\alpha\alpha'},\,\epsilon_{\alpha\alpha'}$ and decomposing $\Lambda^{\alpha\alpha'\dot{\alpha}'}$ as follows: 
\begin{equation}
    \Lambda^{\alpha\alpha'\dot{\alpha}'}_I={\rm i}\,(\gamma^k)^{\alpha\alpha'}\hat{\chi}^{\dot{\alpha}'}_{Ik}+
    \epsilon^{\alpha\alpha'}\chi^{\dot{\alpha}'}_I\,,\label{twist2}
\end{equation}
This twist is suggested by the ansatz (\ref{ansatz}) in which the field $\varPsi_{iI}^\alpha$ is expressed in terms of spin-$1/2$ fields $\chi_{I}^{\dot{\alpha}'}$, to be related to the components on the right-hand side of (\ref{twist2}) in the following.\par
Equations (\ref{twist1}) and (\ref{twist2}) amount to writing the spinors in two different bases. The relation between the corresponding components reads
\begin{align}
\begin{cases}
\chi_I \ = \frac{1}{2}({\rm i}\, \varPsi\!\!\!/_I- \eta_I), \\
\hat{\chi}_{Ii}=\varPsi_{Ii}+\frac{{\rm i}}{2}\,{\gamma_i}\left({\rm i} \, \varPsi\!\!\!/_I + \eta_I\right),
\end{cases}
\hspace{0.5cm}
\begin{cases}
\eta_I \ = -\frac{1}{2} \left[ \ii\, \hat{\chi}\!\!\!/_{I} + \chi_I \right],\\
\varPsi_{iI} = \hat{\chi}_{iI} -\frac{1}{2} \,\gamma_i\,\left(\hat{\chi}\!\!\!/_{I}+ \ii\, \chi_I\right)\,,
\end{cases}
\label{psichipsi}
\end{align}
where the spinor indices have been suppressed.\footnote{{Note that the above manipulations require that the only manifest symmetry acting on the odd sector be the diagonal subgroup ${\rm SL}(2,\mathbb{R})\subset {\rm SL}(2,\mathbb{R})_D\times {\rm SL}(2,\mathbb{R})'_{2}$
 so that the three indices $\alpha,\,\alpha',\, \dot{\alpha}'$ are treated on an equal footing.}}
The spinor equations (\ref{eqpsieta}) in the new fields have the following form: 
\begin{align}
(\boldsymbol{\alpha}+1)\epsilon^{\ell ik}\,\nabla_i\left(\hat\chi_{kI}- \frac{\ii}{2}\,\gamma_k\,(\chi_I-\ii\,\hat{\chi}\!\!\!/_{I})\right)+\frac{1-\boldsymbol{\alpha}}{2}\,\nabla^\ell(\chi_I+\ii\,\hat{\chi}\!\!\!/_{I})&=0\,,
\\
\frac{1}{2}\,\left({-}\ii\,\slashed{\nabla}\chi_I-\frac{2}{L}\chi_I\right)-\frac{\ii}{2}\,\left({-}\ii\,\slashed{\nabla}\hat{\chi}\!\!\!/_I+
\frac{2}{L}\hat{\chi}\!\!\!/_I\right)+\nabla^i \hat\chi_{iI}&=0\,, \label{eqchi}
\end{align}
and the field variations are
\begin{align}\label{susytransfchi}
    \delta_\varepsilon \phi^{\alpha A}_I &= \ii \left( \gamma^k \right)^\alpha_{\phantom{\alpha} \alpha'} \varepsilon^{\alpha' \dot{\alpha}' A} \hat{\chi}_{Ik \dot{\alpha}'} + \varepsilon^{\alpha \dot{\alpha}' A} \chi_{I \dot{\alpha}'} \,, \nonumber \\
    \delta_\varepsilon \hat{\chi}^{\dot{\alpha}'}_{I l} &= \frac{\ii}{8} \nabla_i \phi^{\alpha A}_I \left[ \left( \gamma_l \gamma^i \right)_{\alpha\beta'}\varepsilon^{\beta'\dot\alpha'}_A - \boldsymbol{\alpha} \left( \gamma_l \right)_{\alpha\beta'} \left( \gamma^i \right)^{\dot{\alpha}'}{}_{\dot{\beta}'} \varepsilon^{\beta' \dot{\beta}'}_{ A}\right]  +\frac{\ii}{2} \mathbb{N}^\alpha_{IA} \left( \gamma_l \right)_{\alpha \alpha'} \varepsilon^{\alpha' \dot{\alpha}' A} \,, \nonumber \\
    \delta_\varepsilon \chi^{\dot{\alpha}'}_I &= \frac{1}{8} \nabla_i \phi^{\alpha A}_I \left[ \left( \gamma^i \right)_{\alpha\beta'} \varepsilon^{\beta' \dot{\alpha}'}_A - \boldsymbol{\alpha} \epsilon_{\alpha\beta'} \left( \gamma^i \right)^{\dot{\alpha}'}{}_{\dot{\beta}'}\varepsilon^{\beta' \dot{\beta}'}_{ A} \right]  +\frac{1}{2} \mathbb{N}^\alpha_{IA}\epsilon_{\alpha\alpha'} \varepsilon^{\alpha' \dot{\alpha}' A} \,.
\end{align}

\paragraph{The Lagrangian and the supersymmetry variations\\}
In terms of the fields $\hat{\chi}^{\dot{\alpha}'}_{Ik}$, $\chi^{\dot{\alpha}'}_I$, the spacetime Lagrangian \eqref{sptlag} takes the form
\begin{align}\nonumber
\mathcal L_{\rm spacetime}=&\frac 12 \nabla_{{i}} \phi^{\alpha A}_{I}\, \nabla^{{i}} \phi^{\beta BI} \epsilon_{\alpha\beta}\epsilon_{AB}-\mathcal V(\phi)+\\ \nonumber
&-\frac 4{\boldsymbol{\alpha}^2-1}\left[\ii \epsilon^{ijk} \hat \chi_{i I}^t\epsilon \nabla_j \hat\chi_{k}^I + \boldsymbol{\alpha} \hat \chi_{i I}^t\epsilon \gamma^j\nabla_j \hat\chi_{i}^I+
 \alpha \chi_I^t \epsilon\gamma^i \nabla_i \chi^I-2\ii \chi_I^t\epsilon \nabla^i    \hat\chi_{i}^I\right]+\nonumber\\
&+\frac {2\ii}{L(\boldsymbol{\alpha} + 1)}\left[\hat\chi_{k I}^t \epsilon\hat\chi^{kI}- \ii \epsilon^{ijk}\hat\chi_{i I}^t\epsilon\gamma_j \hat\chi_{k}^I+ \chi_I^t\epsilon\chi^I - 2\ii \hat\chi_{i I}^t\epsilon\gamma^i \chi^I\right] \,, \label{unclag}
\end{align}
where we have used the matrix notation for the bispinor index: $\xi^t\epsilon\zeta\equiv \xi^\alpha\zeta^\beta\,\epsilon_{\alpha\beta}$. \newline
The above results can be further rewritten by decomposing the $\hat{\chi}_{Ii}$ field into its spin-$3/2$ and spin-$1/2$ components, $\mathring{\chi}_{Ii}$ and $\hat{\slashed{\chi}}_{I}$, as follows:
\begin{equation}\label{usefulformula}
    \hat{\chi}_{Ii}=\mathring{\chi}_{Ii}+\frac{1}{3}\,\gamma_i\,\hat{\slashed{\chi}}_{I}\,,
\end{equation}
where $\gamma^i\,\mathring{\chi}_{Ii}=0$. In this way we can rewrite the expressions of $\varPsi_{Ii}$ and $\eta_I$ in \eqref{psichipsi} in the form
\begin{align}
  \varPsi_{Ii}=\mathring{\chi}_{Ii}-\frac{\ii}{2}\,\gamma_i\,\chi_{1I}\,,\qquad \eta_I=-\frac{1}{2}\,\chi_{2I}\,,\label{psietachi}
\end{align}
where $\chi_{1I}$ and $\chi_{2I}$ are defined as follows:
\begin{align}
    \chi_{1I}\equiv -\frac{\ii}{3}\,\hat{\slashed{\chi}}_I+ \chi_I\,,\qquad \chi_{2I}\equiv \ii\,\hat{\slashed{\chi}}_I+ \chi_I\,.
\end{align}
The inverse relations are
\begin{align}
    \chi_{I}= \frac{1}{4}\,\left(3\chi_{1I}+ \chi_{2I}\right) \,, \qquad
     \slashed{\chi}_{I}=\frac{3}{4}\ii\,\left(\chi_{1I}- \chi_{2I}\right)\,,
\end{align}
and the Lagrangian \eqref{unclag}, when expressed in terms of the fields $\mathring{\chi}_{Ii},\,\chi_{1I},\,\chi_{2I}$, takes the simpler form
\begin{align}\label{lagchi0}
    \mathcal L_{\rm spacetime}&=\frac 12 \nabla_{{i}} \phi^{\alpha A}_{I}\, \nabla^{{i}} \phi^{\beta B I} \epsilon_{\alpha\beta}\epsilon_{AB}-\mathcal V(\phi)+ \nonumber\\
    &+\frac{4\ii}{1-\boldsymbol{\alpha}}\,\left[\epsilon^{ijk}\mathring{\chi}_{Ii}^t\epsilon \nabla_j\mathring{\chi}_{k}^I-\frac{\ii}{2}\chi_{1I}^t\epsilon \slashed{\nabla}\,\chi_{1}^I+\mathring{\chi}_{Ii}^t\epsilon\nabla^i\,\chi_{1}^I\right]+\nonumber\\
    &+\frac{2\ii}{\boldsymbol{\alpha}+1}\left[2\,\mathring{\chi}_{Ii}^t\epsilon\,\nabla^i\,\chi_{2}^I+\ii\,\chi_{1I}^t\,\epsilon\,\slashed{\nabla}\chi_{2}^I+\frac{1}{L}\,\chi_{2I}^t\epsilon\,\chi_{2}^I\right] \,,
\end{align}
where for the spinor bilinears we have used the notation illustrated in Appendix \ref{AppA}.

The field equations are readily written in the following form:
\begin{align}
 \delta \mathring{\chi}:& \quad {\mathbb{P}_{\left(\frac{3}{2}\right)\,}}^l{}_i\left((\boldsymbol{\alpha}+1)\,\epsilon^{ijk}\,\nabla_j\left(\mathring{\chi}_{Ik}-\frac{\ii}{2}\,\gamma_k\,\chi_{1I}\right)-\frac{(\boldsymbol{\alpha}-1)}{2}\,\nabla^i\,\chi_{2I}\right)=0 \,, \label{eqcho0}\\ 
 \delta {\chi_1}:& \quad {\nabla}^i\mathring{\chi}_{iI}+ \ii \slashed{\nabla} \chi_{1I} +\ii  \frac{\alpha-1}{2(1+\boldsymbol{\alpha})} \slashed{\nabla} \chi_{2I}=0 \,, \label{eqchi1}\\
\delta {\chi_2}:& \quad -2{\nabla}^i\mathring{\chi}_{iI} {+}\ii \slashed{\nabla} \chi_{1I} +  \frac{2}{L} \chi_{2I}=0 \label{eqchi2}\,,
\end{align}
where ${\mathbb{P}_{\left(\frac{3}{2}\right)}}^j{}_i\equiv {\bf 1}\,\delta_i^j-\frac{1}{3}\,\gamma^j\gamma_i$ is the projector on the spin-$3/2$ representation.
Combining the three equations above, we get the following conditions:
\begin{align}\nonumber
    &{\nabla}^i\mathring{\chi}_{iI}=\frac 16\frac{1-\boldsymbol{\alpha}}{1+\boldsymbol{\alpha}}\left[\ii  \slashed{\nabla} \chi_{2I} - \frac{4(1+\boldsymbol{\alpha})}{L(\boldsymbol{\alpha}-1)}\chi_{2I}\right] \,,\\
    &2\,\epsilon^{ijk}\nabla_j\mathring{\chi}_{Ik}+\left(\nabla^i-\gamma^i \slashed\nabla\right)\,\chi_{1I}-\frac{\boldsymbol{\alpha}-1}{\boldsymbol{\alpha}+1}\,\nabla^i\chi_{2I}=0\,.
\end{align}

\paragraph{Massive Dirac spinors and unconventional supersymmetry\\}
Let us show that the solutions to equations (\ref{eqcho0}), (\ref{eqchi1}) and (\ref{eqchi2}) comprise a massive Dirac spinor. To this end it suffices to restrict to solutions satisfying the condition
\begin{equation}
    \nabla^i\mathring{\chi}_{Ii}=0\,.\label{nablaichii}
\end{equation}
Equations (\ref{eqchi1}) and (\ref{eqchi2}) can then be rewritten in the following equivalent form:
\begin{align}
       \slashed{\nabla}\chi_{1I}&=\frac{2\ii}{L}\,\chi_{2I}\,\,,\label{higher}\\
  \ii\,\slashed{\nabla}\chi_{2I}&=m\,\chi_{2I}\,,    \label{Dirachi}
\end{align}
where
\begin{equation}
m=\frac{4(\boldsymbol{\alpha}+1)}{L(\boldsymbol{\alpha}-1)}\,.\label{massDirachi}
\end{equation}
The fields $\chi_{2I}$ are now massive Dirac spinors of mass $m$.
Note that, as expected, this mass depends on the parameter $g=(\boldsymbol{\alpha} +1)$, that is on the gauging of the R-symmetry ${\rm SU}(2)$.
As for $\chi_{1I}$, it is straightforward to verify that, given the field $\chi_{2I}$, solution of (\ref{Dirachi}), $\chi_{1I}$ is given by  the following general expression:
\begin{equation}
    \chi_{1I}=-\frac{2}{Lm}\,\chi_{2I}+\sigma_I=\frac{1-\boldsymbol{\alpha}}{2(\boldsymbol{\alpha}+1)}\,\chi_{2I}+\sigma_I\,,\label{chichi2sigma}
\end{equation}
where $\sigma_I$ are massless spinor fields: $\ii\,\slashed{\nabla}\sigma_I=0$. {This result is also expected, since  $ \chi_{1I}$ satisfies a higher order field equation (see \eqref{higher}).}\\ 
We could impose on the solutions to equations (\ref{eqcho0}), (\ref{eqchi1}) and (\ref{eqchi2}) a stronger condition and set the spin-$3/2$ field to zero:
\begin{equation}
  \mathring{\chi}_{Ii}=0\,.\label{chii0}
\end{equation}
This allows to make contact with unconventional supersymmetry, where the fields $\varPsi_{I{i}}^\alpha$ have a vanishing spin-$3/2$ component. 
Equations \eqref{higher} and \eqref{Dirachi} still hold. Now, however, $ \chi_{1I},\, \chi_{2I}$ also satisfy the additional equation
\begin{equation}
        \nabla^i\left((\boldsymbol{\alpha}-1)\,\chi_{2I}-(\boldsymbol{\alpha}+1)\,\chi_{1I}\right)=-\frac{2\ii}{L}\,(\boldsymbol{\alpha}+1)\,\gamma^i\,\chi_{2I}\,,\label{eqchi12}
\end{equation}
which implies that the spin-$3/2$ component of the covariant derivative on the left-hand side vanishes.\\
From eqs. (\ref{psietachi}) we recover a generalisation of the ansatz (\ref{ansatz}),
\begin{align}
  \varPsi_{Ii}&=-\frac{\ii}{2}\,\gamma_i\,\chi_{1I}\,\,,\,\,\,\,
  \eta_I=-\frac{1}{2}\,\chi_{2I}\,,\label{psietachi2}
\end{align}
where $ \varPsi_{Ii}$ only has a spin-$1/2$ component $\chi_{1I}$ which is expressed in terms of the  massive spinor field $\chi_{2I}$ through (\ref{chichi2sigma}).
The propagating spinor $\chi^{(\rm{AVZ})}_I$ of the model \cite{Alvarez:2011gd}, appearing in eq. \eqref{ansatz}, has to be identified, using \eqref{psietachi}, with
$$\chi^{(\rm{AVZ})}_I=\ii \gamma^i \varPsi_{iI} =\frac 32 \chi_{1I}\,.$$
We have discussed above only those solutions for which either eq. (\ref{nablaichii}) or the stronger one (\ref{chii0}) holds. Our supersymmetric model, however, features more general solutions which non-trivially involve the spin-$3/2$ fields and whose physical applications deserve investigation. We postpone this analysis to future work.

Finally, we notice that the condition for unconventional supersymmetry $\mathring\chi_{iI}=0$ breaks, in general, all supersymmetries of our superspace. Indeed we can use the supersymmetry variations of $\mathring\chi_{iI}$ in the twisted form to write 
\begin{align}
    \delta_{\varepsilon^B\mathcal S_B}\mathring{\chi}_{iI}&=-\frac{\ii(1+\boldsymbol{\alpha})}{8}(\mathbb P_{ij}\nabla^j\phi^A_I)\varepsilon_A \,, \\ \nonumber
    \delta_{\varepsilon^{lB}\mathcal S_{lB}}\mathring{\chi}_{iI}&=\mathbb P_{\left(\frac{3}{2}\right)ij}\,\delta_{\varepsilon^{lB}\mathcal S_{lB}}\varPsi^j_I\\ \nonumber
    &=\frac{\ii}{8}\left((1-\boldsymbol{\alpha})\epsilon_{ijk}\nabla^j\phi^{ A}_I+\frac{1+\boldsymbol{\alpha}}{L}\phi^{ A}_I\eta_{ik}\right)\varepsilon^k_A+\\
    &-\frac{\ii}{24}\gamma_i\left(3\ii(\boldsymbol{\alpha}-1)\mathbb P_{kj}\nabla^j\phi^A_I-2\ii(\boldsymbol{\alpha}-1)\nabla_k\phi^A_I+\frac{\boldsymbol{\alpha}+1}{L}\gamma_k\phi^A_I\right)\varepsilon^k_A \,,
\end{align}
from which it follows that, in general, the vanishing of $\mathring\chi_{iI}$ is not preserved by supersymmetry transformations in $\mathrm{D}^2(2,1;\,\boldsymbol{\alpha})$ superspace. It is important to emphasise, however, that this supersymmetry is not related to the ``unconventional'' one exhibited by the model in \cite{Alvarez:2011gd}, that originated from a target-space symmetry.

\section{Conclusions and Outlook}
In this final Section, we summarize the outcome of our analysis and we conclude with some comments on future developments and perspectives. 

Our results can be summarised as follows:
\begin{itemize}
    \item[1)] We have constructed a three-dimensional model of rigid supersymmetry featuring eight supercharges on a curved AdS$_3$ worldvolume background whose superspace is based on the supergroup $\mathrm{D}^2(2,1;\,\boldsymbol{\alpha})$. The resulting model describes 
    the dynamics of a set of hypermultiplets $(\Lambda^{\alpha\alpha'\dot{\alpha}'}_I,\phi^{\alpha A}_I)$. 
     
    A peculiarity of the chosen superalgebra is the presence of a parameter $\boldsymbol{\alpha}$, independent of the cosmological constant, which defines, through the combination $g=\boldsymbol{\alpha}+1$, the gauging of an internal SU$(2)$, in the absence of dynamical gauge fields. 
    To clarify the meaning of the word ``gauging'' in the present context, let us notice that the ``coupling constant'' $g$ generates a fermion shift $\mathbb N^{\alpha A}_I$ in the supersymmetry transformation of the hyperini \eqref{susytransf}, together with mass terms for the latter fields and non-trivial scalar dynamics. This feature is not fully apparent from the Lagrangian, 
    since the structure of $\mathrm{D}^2(2,1;\,\boldsymbol{\alpha})$ naturally leads to a redundant (4-component) description of the spinorial degrees of freedom and to a generalised definition of gamma matrices which include a dependence on the parameter $\boldsymbol{\alpha}$. Consequently, the dependence of the Lagrangian on the parameter $\boldsymbol{\alpha}$ is somewhat concealed in the matrices $\mathbb{M}^i_\pm$. 
    The proper definition of the hyperini mass requires a formulation of the above fields as ordinary 2-component spinors, and this in turn implies a choice of the Lorentz symmetry in the superspace. This reformulation is intimately related to the issue of the twists.
    
    \item[2)] Two inequivalent twists have indeed been performed, corresponding to two different  identifications of the Lorentz group.

    In the first one, the Lorentz group is identified with the diagonal subgroup of the two $\mathrm{SL}(2,\mathbb{R})$ factors in the AdS$_3$ isometry group. This is the counterpart, in our setting, of the topological twist discussed in \cite{Rozansky:1996bq, Kapustin:2009cd}. After performing this twist, the hyperini decompose into an abelian gauge connection $\varPsi^\alpha_{iI}$ associated with odd symmetry generators, transforming as a vector with respect to the Lorentz group and in a Grassmann-valued field, $\eta^\alpha_I$,  singlet of the Lorentz group (see eq. \eqref{twist1}). Correspondingly, the supersymmetry generators decompose into  vector-like and scalar-like odd generators $\mathcal{S}^A_i, \mathcal{S}^A$. 
    Our spacetime Lagrangian takes the form of a Chern-Simons Lagrangian for  $\varPsi^\alpha_{iI}$, plus a term in the image of $\mathcal{S}^A$, containing the interaction with the other fields.

    An alternative twist can  be performed, involving the SL$(2,\mathbb R)$ flavour group. More precisely, in this case the Lorentz group is defined to be the diagonal of the previously chosen Lorentz group with the flavour SL$(2,\mathbb R)$ acting on the index $\alpha$ carried by the dynamical fields of the model. This corresponds to identifying the Lorentz group as the diagonal of the three groups $\mathrm{SL}(2,\mathbb R)'_1$, $\mathrm{SL}(2,\mathbb R)'_2$, $\mathrm{SL}(2,\mathbb R)_{\mathrm{flavour}}$.
    In particular, the anticommuting fields $\varPsi^\alpha_{i I}$ and $\eta^\alpha_I$ now transform in half-integer Lorentz representations, which are appropriate to their spin statistics, while $\phi^{\alpha A}_I$ transform as commuting spin-$1/2$ fields, fully decoupled from the rest.
    In this new setting, the superspace structure is not manifest.
    This identification, which in fact describes a subsector of the first twist, unveils the interesting structure described by the Lagrangian \eqref{lagchi0}.
    Indeed,  $\varPsi^\alpha_{i I}$ and $\eta^\alpha_I$ acquire a natural  interpretation as spinorial fields, in particular as a purely spin-$3/2$ field $\mathring {\chi}_{iI}$ coupled to two spin-$1/2$ particles $\chi_{1I}$, $\chi_{2I}$. A subset of the solutions of the field equations, defined by the condition $\nabla_i\mathring\chi^i_I=0$,
    describes a massive spin-1/2 field, $\chi_{2I}$, which acts as a source for the field $\chi_{1I}$. The latter can therefore be expressed  as a combination of $\chi_{2I}$ and an arbitrary massless spin-$1/2$ field.\\
    Imposing the stronger condition $\mathring{\chi}_{iI}=0$, we recover the most general solution of the model with unconventional supersymmetry of \cite{Alvarez:2011gd,Andrianopoli:2018ymh,Andrianopoli:2019sip}. It is worth emphasizing, however, that the complete set of solutions of the field equations of our model is richer and describes non-trivial dynamics involving $\mathring {\chi}_{iI}$, $\chi_{1I}$, $\chi_{2I}$, yet to be explored.
\end{itemize}
A few questions, however, remain open. We have found that the action, though perfectly supersymmetric in spacetime, is only quasi-supersymmetric when extended as a three-form in the full superspace, its invariance requiring the addition of boundary terms. This is possibly related to the presence, for $\boldsymbol{\alpha}+1\neq 0$, of an ${\rm SU}(2)$-gauging involving non-dynamical vector fields $A^x_\mu$ which, in our model, are frozen as background fields. As $\boldsymbol{\alpha}+1$ is set to zero, indeed, the full supersymmetry of the Lagrangian is restored in superspace.\\
In this same limit the interpretation of the $\mathcal{S}^A$ generators as anticommuting BRST operators is recovered. \\
As $\boldsymbol{\alpha}$ is set to this singular value, the fermion masses, which would vanish in the model considered here, could instead be obtained through the gauging of the flavour group $\mathcal{G}={\rm SL}(2,\mathbb{R})\times {\rm SO}(2)$, according to the analysis of \cite{Andrianopoli:2019sqe}. This gauging, which we did not consider in the present work, can be regarded as more \emph{conventional} in that it will involve gauge fields sitting in vector multiplets and thus that are not background fields. The flavour group, however, can also be gauged for a generic value of $\boldsymbol{\alpha}$. \par
Let us now turn to the discussion of perspectives and future developments. Given the peculiar structure of the supergroup considered here and the chosen dynamical supermultiplets, there are multiple possible routes.\\
A first choice would be, as mentioned above,  to introduce a proper gauging of the flavour symmetry group: we expect this idea to lead to a structure similar to the one of \cite{Kapustin:2009cd}, where the spacetime Lagrangian truly is a CS theory, having both even and odd connections. Further insight could be found by including in our analysis also the interaction with a set of twisted hypermultiplets \cite{Koh:2009um}. It would then be interesting to explore the possible embedding of the full theory on $\mathrm{D}^2(2,1;\,\boldsymbol{\alpha})$ superspace in higher-dimensional supergravity within a holographic setting. \\
Another possibility for extending the present analysis is given by the choice of a more general, curved scalar manifold of hyper-K\"ahler type. In this case the hypermultiplet Lagrangian should be modified by the addition of terms accounting for the curvature of the hyper-K\"ahler geometry, as sketched in Section \ref{hyperkahlerstructure}.\\
Finally, it would be appealing to consider the $\mathrm D^2 (2,1;\,\boldsymbol{\alpha})$ superalgebra as a framework to derive new models of interacting massive Dirac particles, whose application to the description, for instance, of graphene-like materials \cite{Iorio:2018agc}, similarly to what has been done for models with unconventional supersymmetry, is an interesting task to be pursued.

\section*{Acknowledgments}

We thank P.A. Grassi and R. D'Auria for useful suggestions and interesting discussions. \\
L.R. would like to thank the Department of Applied Science and Technology of the Polytechnic of Turin for financial support.

\appendix
\section{Useful Relations}\label{AppA}
In this Appendix we state some of the properties of the Clifford algebra and of the matrices used throughout the text. We are particularly interested in the SL$(2,\mathbb R)$ factors, appearing both in the bosonic subalgebra of $\mathrm{D}^2(2,1;\,\boldsymbol{\alpha})$ and as a flavour group and in their interplay.\\
For this reason, in this Appendix we will not distinguish between different types of spinorial indices (e.g. $\alpha,\alpha',\dot\alpha'$), unless explicitly stated and we will identify spacetime indices belonging to different SL$(2,\mathbb R)$ factors, since we take the diagonal group SL$(2,\mathbb R)_D$ as Lorentz symmetry of our theory.\\
We adopt the following general conventions on gamma matrices: 
\begin{align}
    \{\gamma^i,\gamma^j\}=2\eta^{ij}\mathbb{1}_{2\times2}\,, \quad [\gamma^i,\gamma^j]\equiv 2\gamma^{ij}=2\ii\epsilon^{ijk}\gamma_k\,,
\end{align}
Once the SL$(2,\mathbb R)$-invariant tensor $\epsilon_{12}=\epsilon^{12}=1$ is introduced, one can lower and raise the indices of the gamma matrices in the following way:
\begin{align}
    ( \gamma^i )_{\alpha \beta} = \epsilon_{\alpha \gamma} ( \gamma^i )^\gamma_{\phantom{\gamma}\beta},\qquad ( \gamma^i )^{\alpha \beta} = ( \gamma^i )^\alpha_{\phantom{\alpha}\gamma} \epsilon^{\gamma \beta} \,,
\end{align}
where the obtained matrices are symmetric.\\
The antisimmetric matrix $\epsilon_{\alpha\beta}$ satisfies
$$\epsilon^{\alpha\beta}\epsilon_{\rho\sigma}=\delta^\alpha_\rho\delta^\beta_\sigma-\delta^\alpha_\sigma\delta^\beta_\rho\,, \qquad \epsilon^{\alpha\beta}\epsilon_{\beta\gamma}=-\delta^\alpha_\gamma\,,$$ whereas the sum of all gamma matrices with uncontracted indices yields
$$ (\gamma^i)^{\alpha \beta} ( \gamma_i )_{\rho \sigma} = - (\delta^\alpha_\rho\delta^\beta_\sigma+\delta^\alpha_\sigma\delta^\beta_\rho)\,.$$
The conventions used in the text for traces and spinor bilinears are the following:
\begin{align}
    \mathrm{Tr}(\gamma^i\gamma^j)\equiv(\gamma^i)^\alpha{}_\beta(\gamma^j)^\beta{}_\alpha\,, \quad \lambda^t\epsilon\chi\equiv\lambda^\alpha\epsilon_{\alpha\beta}\chi^\beta\,, \quad \lambda^t\epsilon\gamma^i\chi\equiv\lambda^\alpha\epsilon_{\alpha\beta}(\gamma^i)^{\beta}{}_\gamma\chi^\gamma\,, \quad (\gamma_i)^t= \epsilon (\gamma_i)\epsilon \,,
\end{align}
where $\lambda,\chi$ are two generic spinors and the upper $t$ denotes transposition.\\
Other conventions, needed to justify the form of the $\mathrm{D}^2(2,1;\,\boldsymbol{\alpha})$ superalgebra, concern the properties of spinors and spinorial forms under complex conjugations,
\begin{equation}
    \left( \lambda^\alpha \psi^\beta \right)^* \equiv \psi^{\beta *} \lambda^{\alpha *} \,,
\end{equation}
\begin{equation}
    \left( \diff \theta^\alpha\wedge \diff \theta^\beta \right)^* = - \diff \theta^\alpha \wedge\diff \theta^\beta \,.
\end{equation}
We end up with a list of the properties of the $ 4\times4$-matrices $\left(\mathbb T^i_{(1)}\right)^{(\alpha)}{}_{(\beta)}$ and $\left(\mathbb T^i_{(2)}\right)^{(\alpha)}{}_{(\beta)}$ defined in the main text, where $(\alpha)=\alpha'\dot\alpha'$:
\begin{align}\nonumber
\left( \mathbb{T}^i_{(1)} \right)^{(\alpha)}_{\phantom{(\alpha)} (\beta)} \left( \mathbb{T}^j_{(1)} \right)^{(\beta)}_{\phantom{(\beta)} (\gamma)} \,&=\, -\frac 14 \eta^{ij} \delta^{(\alpha)}_{(\gamma)}- \frac 12 \epsilon^{ijk} \left( \mathbb{T}_{(1)k} \right)^{(\alpha)}_{\phantom{(\alpha)} (\gamma)}\,,\\ \nonumber
\left( \mathbb{T}^i_{(2)} \right)^{(\alpha)}_{\phantom{(\alpha)} (\beta)} \left( \mathbb{T}^j_{(2)} \right)^{(\beta)}_{\phantom{(\beta)} (\gamma)} \,&=\, -\frac 14 \eta^{ij} \delta^{(\alpha)}_{(\gamma)}- \frac 12 \epsilon^{ijk} \left( \mathbb{T}_{(2)k} \right)^{(\alpha)}_{\phantom{(\alpha)} (\gamma)}\,,\\ \nonumber
\left( \mathbb{T}^i_{(1)} \right)^{(\alpha)}_{\phantom{(\alpha)} (\beta)} \left( \mathbb{T}^j_{(2)} \right)^{(\beta)}_{\phantom{(\beta)} (\gamma)} \,&=\ \left( \mathbb{T}^j_{(2)} \right)^{(\alpha)}_{\phantom{(\alpha)} (\beta)} \left( \mathbb{T}^i_{(1)} \right)^{(\beta)}_{\phantom{(\beta)} (\gamma)} \,
=\, -\frac 14 \left( \gamma^i \right)^{\alpha'}_{\phantom{\alpha'} \gamma'}  \otimes \left( \gamma^j \right)^{\dot\alpha'}_{\phantom{\dot\alpha'} \dot\gamma'}\,.
\\
\sum_{a=1}^2(\mathbb{T}_{(a)}^i)^{(\alpha)(\beta)}(\mathbb{T}_{(a)\,i})_{(\rho)(\sigma)}&=\delta^{(\alpha)}_{(\rho)}\delta^{(\beta)}_{(\sigma)}-\delta^{(\alpha)}_{(\sigma)}\delta^{(\beta)}_{(\rho)}\,,\nonumber\\
(\mathbb{T}_{(1)}^i\mathbb{T}_{(2)}^j)^{(\alpha)(\beta)}(\mathbb{T}_{(1)\,i}\mathbb{T}_{(2)\,j})_{(\rho)(\sigma)}&=\frac{1}{8}\left(\delta^{(\alpha)}_{(\rho)}\delta^{(\beta)}_{(\sigma)}+\delta^{(\alpha)}_{(\sigma)}\delta^{(\beta)}_{(\rho)}\right)\,-\frac{1}{16}\,\delta^{(\alpha)(\beta)}\,\delta_{(\rho)(\sigma)}\,,
\end{align}
from which we see that
\begin{equation}
    \mathrm{Tr} \left( \mathbb{T}^i_{(1)} \mathbb{T}^j_{(1)} \right)= \mathrm{Tr} \left( \mathbb{T}^i_{(2)} \mathbb{T}^j_{(2)} \right)= - \eta^{ij}\,; \quad  \mathrm{Tr}\left( \mathbb{T}^k_{(1)} \mathbb{T}^i_{(2)} \right) = \mathrm{Tr}\left( \mathbb{T}^k_{(2)} \mathbb{T}^i_{(1)} \right) = 0 \,.
\end{equation}

\clearpage



\begin{thebibliography}{99}

\bibitem{Witten:1988hf}
E.~Witten,
``Quantum Field Theory and the Jones Polynomial,''
Commun. Math. Phys. \textbf{121} (1989), 351-399
doi:10.1007/BF01217730

\bibitem{Rozansky:1996bq}
L.~Rozansky and E.~Witten,
``HyperKahler geometry and invariants of three manifolds,''
Selecta Math. \textbf{3} (1997), 401-458
doi:10.1007/s000290050016
[arXiv:hep-th/9612216 [hep-th]].

\bibitem{Gaiotto:2008sd}
 D.~Gaiotto and E.~Witten, \\
 ``Janus Configurations, Chern-Simons Couplings, And The theta-Angle in N=4 Super Yang-Mills Theory,''
 JHEP {\bf 1006} (2010) 097
 doi:10.1007/JHEP06(2010)097
 [arXiv:0804.2907 [hep-th]].

\bibitem{Kapustin:2009cd}
  A.~Kapustin and N.~Saulina,
  ``Chern-Simons-Rozansky-Witten topological field theory,''
  Nucl.\ Phys.\ B {\bf 823} (2009) 403
  doi:10.1016/j.nuclphysb.2009.07.006
  [arXiv:0904.1447 [hep-th]].

\bibitem{Koh:2009um}
E.~Koh, S.~Lee and S.~Lee,
``Topological Chern-Simons Sigma Model,''
JHEP \textbf{09} (2009), 122
doi:10.1088/1126-6708/2009/09/122
[arXiv:0907.1641 [hep-th]].

\bibitem{Alvarez:2011gd}
  P.~D.~Alvarez, M.~Valenzuela and J.~Zanelli,
 ``Supersymmetry of a different kind,''
  JHEP {\bf 1204} (2012) 058
  doi:10.1007/JHEP04(2012)058
  [arXiv:1109.3944 [hep-th]].

\bibitem{Andrianopoli:2018ymh}
L.~Andrianopoli, B.~L.~Cerchiai, R.~D'Auria and M.~Trigiante,
``Unconventional supersymmetry at the boundary of AdS$_{4}$ supergravity,''
JHEP \textbf{04} (2018), 007
doi:10.1007/JHEP04(2018)007
[arXiv:1801.08081 [hep-th]].

\bibitem{Andrianopoli:2019sip}
L.~Andrianopoli, B.~L.~Cerchiai, R.~D'Auria, A.~Gallerati, R.~Noris, M.~Trigiante and J.~Zanelli,
``$\mathcal{N}$-extended $D = 4$ supergravity, unconventional SUSY and graphene,''
JHEP \textbf{01} (2020), 084
doi:10.1007/JHEP01(2020)084
[arXiv:1910.03508 [hep-th]].

\bibitem{Andrianopoli:2020zbl}
L.~Andrianopoli, B.~L.~Cerchiai, R.~Matrecano, O.~Miskovic, R.~Noris, R.~Olea, L.~Ravera and M.~Trigiante,
``$ \mathcal{N} $ = 2 AdS$_{4}$ supergravity, holography and Ward identities,''
JHEP \textbf{02} (2021), 141
doi:10.1007/JHEP02(2021)141
[arXiv:2010.02119 [hep-th]].

\bibitem{Achucarro:1987vz}
A.~Achucarro and P.~K.~Townsend,
``A Chern-Simons Action for Three-Dimensional anti-De Sitter Supergravity Theories,''
Phys. Lett. B \textbf{180} (1986), 89
doi:10.1016/0370-2693(86)90140-1

\bibitem{Andrianopoli:2019sqe}
L.~Andrianopoli, B.~L.~Cerchiai, P.~A.~Grassi and M.~Trigiante,
``The Quantum Theory of Chern-Simons Supergravity,''
JHEP \textbf{06} (2019), 036
doi:10.1007/JHEP06(2019)036
[arXiv:1903.04431 [hep-th]].

\bibitem{Delduc:1990je}
  F.~Delduc, C.~Lucchesi, O.~Piguet and S.~P.~Sorella,
  ``Exact Scale Invariance of the {Chern-Simons} Theory in the Landau Gauge,''
  Nucl.\ Phys.\ B {\bf 346}, 313 (1990).
  doi:10.1016/0550-3213(90)90283-J

\bibitem{Vilar:1997fg}
  L.~C.~Q.~Vilar, O.~S.~Ventura, C.~A.~G.~Sasaki and S.~P.~Sorella,
  ``Algebraic characterization of vector supersymmetry in topological field theories,''
  J.\ Math.\ Phys.\  {\bf 39}, 848 (1998)
  doi:10.1063/1.532356
  [hep-th/9706133].

\bibitem{DelCima:1998ux}
  O.~M.~Del Cima, K.~Landsteiner and M.~Schweda,
  ``Twisted N=4 SUSY algebra in topological models of Schwarz type,''
  Phys.\ Lett.\ B {\bf 439}, 289 (1998)
  doi:10.1016/S0370-2693(98)01065-X
  [hep-th/9806137].

\bibitem{Gieres:2000pv}
  F.~Gieres, J.~Grimstrup, T.~Pisar and M.~Schweda,
  ``Vector supersymmetry in topological field theories,''
  JHEP {\bf 0006}, 018 (2000)
  doi:10.1088/1126-6708/2000/06/018
  [hep-th/0002167].

\bibitem{Gunaydin:1986fe}
M.~Gunaydin, G.~Sierra and P.~K.~Townsend,
``The Unitary Supermultiplets of $d=3$ Anti-de Sitter and $d=2$ Conformal Superalgebras,''
Nucl. Phys. B \textbf{274} (1986), 429-447
doi:10.1016/0550-3213(86)90293-2

\bibitem{Festuccia:2011ws}
G.~Festuccia and N.~Seiberg,
``Rigid Supersymmetric Theories in Curved Superspace,''
JHEP \textbf{06} (2011), 114
doi:10.1007/JHEP06(2011)114
[arXiv:1105.0689 [hep-th]].

\bibitem{Butter:2011zt}
D.~Butter and S.~M.~Kuzenko,
``N=2 supersymmetric sigma-models in AdS,''
Phys. Lett. B \textbf{703} (2011), 620-626
doi:10.1016/j.physletb.2011.08.043
[arXiv:1105.3111 [hep-th]].

\bibitem{Butter:2012vm}
  D.~Butter, S.~M.~Kuzenko and G.~Tartaglino-Mazzucchelli,
 ``Nonlinear sigma models with AdS supersymmetry in three dimensions,''
  JHEP {\bf 1302} (2013) 121
  doi:10.1007/JHEP02(2013)121
  [arXiv:1210.5906 [hep-th]].

\bibitem{DAuria:2020guc}
R.~D'Auria,
``Geometric supergravity,''
[arXiv:2005.13593 [hep-th]].

\bibitem{Frappat:1996pb}
L.~Frappat, P.~Sorba and A.~Sciarrino,
``Dictionary on Lie superalgebras,''
[arXiv:hep-th/9607161 [hep-th]].


\bibitem{TheFormula}
Rossmann, Wulf (2002), ``Lie Groups ? An Introduction Through Linear Groups'', Oxford Graduate Texts in Mathematics, Oxford Science Publications, ISBN 0-19-859683-9.

\bibitem{Breitenlohner:1982bm}
P.~Breitenlohner and D.~Z.~Freedman,
``Positive Energy in anti-De Sitter Backgrounds and Gauged Extended Supergravity,''
Phys. Lett. B \textbf{115} (1982), 197-201
doi:10.1016/0370-2693(82)90643-8

\bibitem{Breitenlohner:1982jf}
P.~Breitenlohner and D.~Z.~Freedman,
``Stability in Gauged Extended Supergravity,''
Annals Phys. \textbf{144} (1982), 249
doi:10.1016/0003-4916(82)90116-6

\bibitem{Andrianopoli:2014aqa}
L.~Andrianopoli and R.~D'Auria,
``N=1 and N=2 pure supergravities on a manifold with boundary,''
JHEP \textbf{08} (2014), 012
doi:10.1007/JHEP08(2014)012
[arXiv:1405.2010 [hep-th]].

\bibitem{Ferrara:1985gj}
  S.~Ferrara and L.~Maiani,
  ``An Introduction To Supersymmetry Breaking In Extended Supergravity,''\\
  published in the proceeding of the 5th SILARG Symposium on Relativity, Supersymmetry and Cosmology, 4-11 Jan 1985. Bariloche, Argentina, CERN-TH-4232/85.

\bibitem{Cecotti:1985mx}
  S.~Cecotti, L.~Girardello and M.~Porrati,
  ``Ward Identities Of Local Supersymmetry And Spontaneous Breaking Of Extended Supergravity,''\\
  published in the proceedings of the
Johns Hopkins Workshop on Current Problems in Particle Theory 9: New Trends in Particle Theory,
5-7 Jun 1985. Florence, Italy, CERN-TH-4256/85.

\bibitem{Andrianopoli:2015rpa}
L.~Andrianopoli, P.~Concha, R.~D'Auria, E.~Rodriguez and M.~Trigiante,
``Observations on BI from $\mathcal{N}=2$ Supergravity and the General Ward Identity,''
JHEP \textbf{11} (2015), 061
doi:10.1007/JHEP11(2015)061
[arXiv:1508.01474 [hep-th]].

\bibitem{Hitchin:1986ea}
  N.~J.~Hitchin, A.~Karlhede, U.~Lindstrom and M.~Rocek,
``Hyperkahler Metrics and Supersymmetry,''
  Commun.\ Math.\ Phys.\  {\bf 108} (1987) 535.

\bibitem{Galicki:1986ja}
  K.~Galicki,
``A Generalization of the Momentum Mapping Construction for Quaternionic Kahler Manifolds,''
  Commun.\ Math.\ Phys.\  {\bf 108} (1987) 117.

\bibitem{Iorio:2018agc}
A.~Iorio and P.~Pais,
``(Anti-)de Sitter, Poincar\'e, Super symmetries, and the two Dirac points of graphene,''
Annals Phys. \textbf{398} (2018), 265-286
doi:10.1016/j.aop.2018.09.011
[arXiv:1807.08764 [hep-th]].


\end{thebibliography}
\end{document}